# Ultrastrong Light-Matter Coupling in

# 2D Metal-Organic Chalcogenates

Surendra B. Anantharaman[1,*,†], Jason Lynch[1,†], Mariya Aleksich[2,3], Christopher E. Stevens[4,5], Christopher Munley[6], Bongjun Choi[1], Sridhar Shenoy[1], Thomas Darlington[7], Arka Majumdar[6,8], P. James Shuck[7], Joshua Hendrickson[5], J. Nathan Hohman[2,3], Deep Jariwala[1,*]

[1] Department of Electrical and Systems Engineering, University of Pennsylvania, Philadelphia, Pennsylvania 19104, United States

[2] Institute of Materials Science, University of Connecticut, Storrs CT, 06269

[3] Department of Chemistry, University of Connecticut, Storrs, CT 06269

[4] KBR Inc., Beavercreek, Ohio 45431, United States

[5] Air Force Research Laboratory, Sensors Directorate, Wright-Patterson Air Force Base, Ohio 45433, United States

[6] Department of Physics, University of Washington, Seattle, Washington 98195, United States

[7] Department of Mechanical Engineering, Columbia University, New York, New York 10027, United States

[8] Department of Electrical and Computer Engineering, University of Washington, Seattle, Washington 98195, United States

[*] Corresponding authors: sba@iitm.ac.in, dmj@seas.upenn.edu

[†] These authors contributed equally to this work.

## Abstract

**Hybridization of excitons with photons to form hybrid quasiparticles, exciton-polaritons (EPs), has been widely investigated in a range of semiconductor material systems coupled to photonic cavities. Self-hybridization occurs when the semiconductor itself can serve as the photonic cavity medium resulting in strongly-coupled EPs with Rabi splitting energies (ħΩ) > 200 meV at room temperatures which recently were observed in layered two-dimensional (2D) excitonic materials. Here, we report an extreme version of this phenomenon, an ultrastrong EP coupling, in a nascent, 2D excitonic system, the metal organic chalcogenate (MOCHA) compound named mithrene. The resulting self-hybridized EPs in mithrene crystals placed on Au substrates show Rabi Splitting in the ultrastrong coupling range (ħΩ > 600 meV) due to the strong oscillator strength of the excitons concurrent with the large refractive indices of mithrene. We further show bright EP emission at room temperature as well as EP dispersions at low-temperatures. Importantly, we find lower EP emission**

**linewidth narrowing to ~1 nm when mithrene crystals are placed in closed Fabry-Perot cavities. Our results suggest that MOCHA materials are ideal for polaritonics in the deep green-blue part of the spectrum where strong excitonic materials with large optical constants are notably scarce.**

## Introduction

Exciton-polaritons are part-light, part-matter quasiparticles that are the result of energy being exchanged between a photon trapped in a cavity and an exciton (Coulomb bound electron-hole pair) state that fundamentally change the optical dispersion of a system. Since exciton-polaritons are the result of strong light-matter interactions, their properties can be leveraged in optoelectronic devices such as lasers[1,2], light-emitting diodes (LEDs)[3,4], and photovoltaics[5,6]. The rate at which energy is exchanged between the light and matter states is described by the coupling parameter (g). When g is smaller than the loss rates of the unperturbed exciton ($\Gamma_X$) and cavity ($\Gamma_C$), the energy is dissipated faster than it is exchanged between the states and no exciton-polaritons are formed. In this case, the system is said to be in the weak coupling regime. However, when the coupling parameter is larger than either state's decay rate ($g > |\Gamma_X - \Gamma_c|/4$), the system enters the strong coupling (SC) regime where the excited light and matter states hybridize to form upper and lower exciton-polaritons with properties of both light and matter[7,8]. In the strong coupling regime, only first-order (absorption and emission) and second-order (scattering) effects need to be considered. However, as the coupling parameter further increases, the exciton-polariton enters the ultrastrong coupling (USC) regime where higher-order effects cannot be ignored[9–11]. The transition between the strong coupling and ultrastrong coupling regimes is continuous unlike the transition between the weak coupling and strong coupling regimes with the sudden hybridization of states, but by convention, the exciton-polariton is said to be in the ultrastrong coupling regime when the coupling parameter is more than 10% of the exciton energy ($E_X$). In the USC regime, a population of virtual photons forms in the ground state of the system shifting its energy. The USC regime enables both exotic quantum mechanical phenomena such as the dynamical Casimir effect[12,13] and photon pair production[14] as well as practical phenomena such as switching on the scale of 10 fs[15]. To date, USC has only been observed in the visible range using closed cavity geometries[16–18], but it has not been observed in a self-hybridized system due to the lack of large band-gap semiconductor materials with strong oscillator strengths in the excitonic resonance.

Self-hybridized exciton-polaritons from transition metal dichalcogenides (TMDCs) have been explored recently for both basic[19] and applied research such as light emitting diodes[20,21]. One of the major limitations in the self-hybridized TMDC is the absence of polariton emission from multilayer TMDCs owing to their indirect bandgap nature. Alternative strategies such as the preparation of TMDC superlattices that decouple electronic interactions in multilayers are yet to demonstrate polariton emission[22]. The largest bandgap material available in the conventional TMDC family ($WS_2$) is around 2 eV, which illustrates the lack of materials with blue emission (~2.5 eV). So far, 3D inorganic materials such as ZnO[23] and GaN[24] are the only materials with polariton emission in the blue region. Further, 3D materials are challenging towards device integration due to lattice strain and complexity involved in the device fabrication. In this work, we report polariton emission in the blue region from a new 2D material – mithrene, which is part of a class of layered, bulk metal-organic chalcogenate materials[25]. Mithrene consists of two-dimensional, inorganic AgSe layers separated by organic insulator layers (phenyl groups) to form a multi-quantum well system[26]. Its excitonic properties make it a strong candidate for optoelectronic devices since it is both direct band gap and supports excitons with binding energies up to 400 meV[27]. High exciton binding energy, along with the advantageous direct band gap of mithrene in the blue region, can be used in light-emitting devices and photodetectors. In addition, the solution-processibility of mithrene at low temperatures (< 200 $^{o}$C) and its van der Waals nature enables easier integration on Si-based platforms compared to its counterparts such as III-V and oxide-based semiconductors where lattice strain is detrimental to device performance. The large exciton binding energy is the result of the quantum-confinement effects of the multi-quantum well geometry, and it enables the excitons to have large oscillator strengths (f) at room temperature. Not only do large oscillator strengths make the excitons highly absorptive, it also increases the coupling parameter of exciton-polaritons as $g \propto \sqrt{\frac{f}{V_m}}$ where $V_m$ is the mode volume of the cavity. Strong excitons in other quantum-confined semiconductors have been shown to be excellent for optoelectronic applications[28,29]. However, mithrene is still a relatively new material so most research has focused on its growth/characterization[26,30–32] or basic optical properties[27].

In this paper, we present the first study of light-matter interactions in mithrene and observe the formation of self-hybridized, ultrastrong coupled exciton-polaritons in mithrene with the largest observed normalized coupling parameter ($g/E_x$ = 0.14) in the visible range. We also find that the light-matter states are emissive allowing for geometrically tunable emission from the exciton wavelength to the lower exciton-polariton wavelength (468 nm to 515 nm). When encapsulated by Ag to form a closed-cavity system, we observed longer exciton-polariton lifetime compared to open-cavity as well as multiple states with ≈1 nm linewidth emission at low temperatures (80 K). Our findings demonstrate that mithrene is a new material that can be used to probe ultrastrong light-matter interactions. The emissive properties of the exciton-polaritons also demonstrates that mithrene can be used as a monochromatic light source in the blue to green region of the visible spectrum.

## Results and Discussion

Our cavity mode is a lossy Fabry-Perot cavity which is formed by the highly reflective substrate and the air-mithrene interface which is reflective due to the large refractive index of mithrene[33,34]. Since the Fabry-Perot cavity is formed by the top and bottom of mithrene crystal, the cavity energy ($E_c$) can be tuned by varying its thickness ($t = \lambda_c/4n$ where t is the thickness of mithrene, $\lambda_c$ is the cavity wavelength, and n is the real part of the refractive index of mithrene). The lossy Fabry-Perot cavity then hybridizes with mithrene's exciton to form a higher energy upper exciton-polariton (UEP) and smaller energy lower exciton-polariton (LEP) where the lower polariton state is found to emit blue light (Figure 1a). The mithrene on gold sample was prepared by drop-casting mithrene in a propylamine-$H_2O$ solution on the Au substrate[26]. Atomic force microscopy (AFM) was used to show that this process yielded mithrene flakes of ~30 µm x 30 µm in lateral area with a thickness of 554 nm (Figure 1b). The refractive index of mithrene is measured using spectroscopic ellipsometry (Supporting Information Figure S1). The reflectance of the system is then simulated using the complex refractive index and the transfer matrix method (TMM)[35]. The TMM is found to accurately predict the energies of the UEP and LEP, and simulations clearly show the anti-crossing behavior for mithrene on Au which is a clear signature of the SC regime (Figure 1c). The coupling parameter is then calculated to be 378 meV by extracting the UEP and LEP energies for various cavity energies and fitting it to the quantum Rabi model[36–38] (Supporting Information Section 1). Here, g is 14% of the exciton energy placing the exciton-polaritons in the USC regime. Most other low-dimensional semiconductors only host exciton-polaritons in the SC regime (Figure 1d). The USC regime is typically achieved by either using low energy transitions in the mid-infrared to microwave ranges which lowers the required coupling parameter and allows for smaller mode volumes relative to the wavelength of light[39], or by coupling a high-Q cavity to organic molecules whose Frankel excitons have extremely large oscillator strengths[40]. However, mithrene differs from these because its band gap is in the blue region of light (468 nm), and the estimated value of its exciton binding energy (400 meV) along with the semiconductor layers being inorganic suggests that it hosts Wannier-Mott excitons[41]. Despite mithrene not using either of these strategies for USC, it still supports exciton-polaritons in the USC regime due to its excitons having extraordinarily large oscillator strengths and the crystal (exciton medium) concurrently possessing a large refractive index (due to its part inorganic nature) that enables small mode volumes.

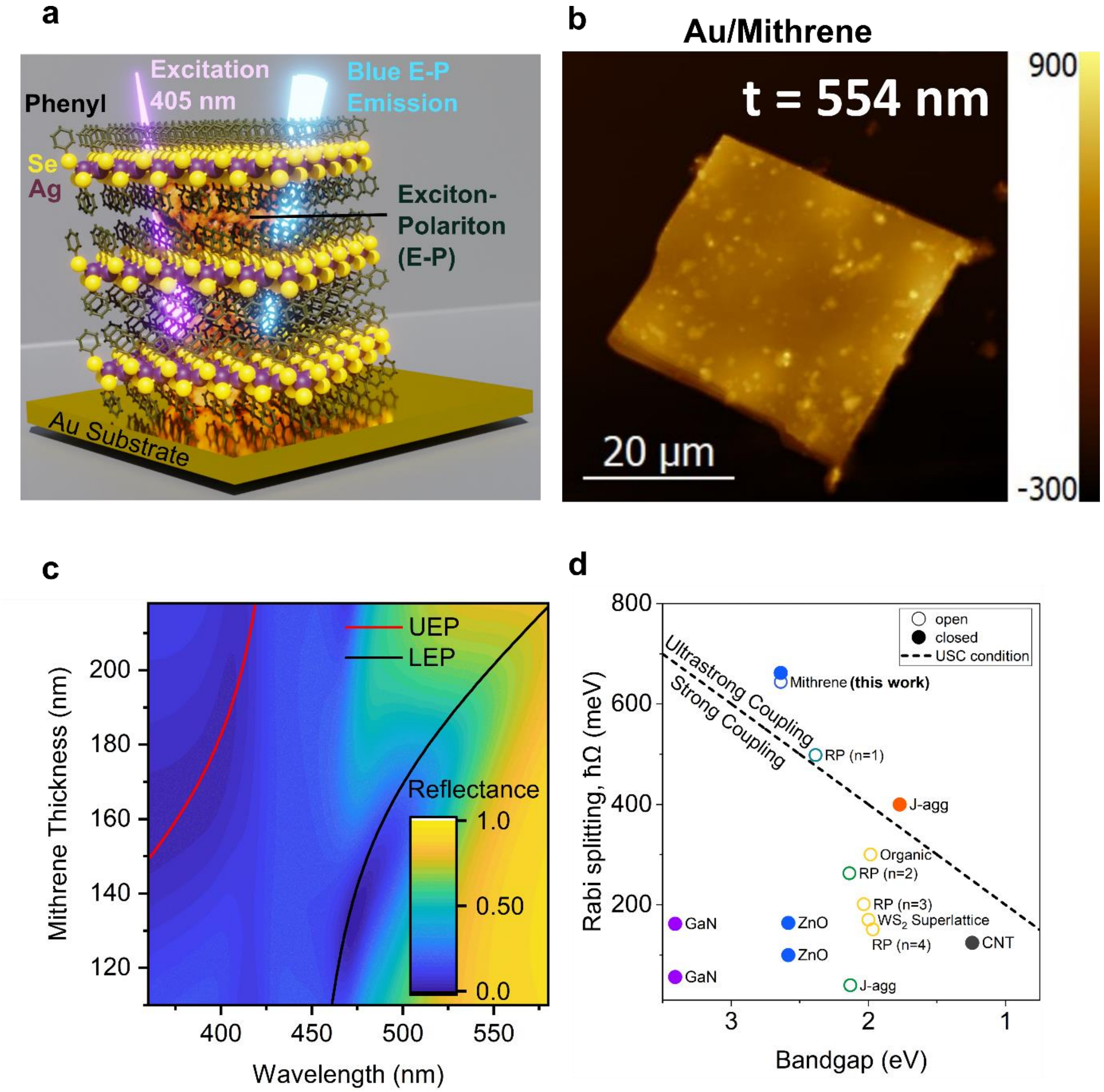

**Figure 1. Self-Hybridized Exciton-Polaritons in Mithrene on Au Substrates.** (a) Schematic of exciton-polariton formation in mithrene crystals placed on the 100 nm Au substrate. (b) AFM image of the drop-casted mithrene from propylamine-$H_2O$ solution on the Au substrate. The thickness of the crystal was identified to be 554 nm, which can support exciton-polaritons in open cavity. (c) transfer-matrix method calculation showing exciton-polariton formation from mithrene on an Au substrate. (d) Comparison of the Rabi splitting of open and closed cavity mithrene with other low-dimensional semiconductors taken from literature (GaN[42], ZnO[43,44], RP (n=1 to 4)[45], $WS_2$ Superlattice[46], organic[47], J-aggregate[48,49], CNT[50]).

To confirm the hybrid nature of our observed absorption modes, thin films of mithrene on an Au substrate are studied since exciton-polaritons cannot form in films much thinner than the wavelength of light ($t < \lambda/4n$). Therefore, the uncoupled exciton states can be seen in a thin film of mithrene (t = 50 nm) using its reflectance and photoluminescence (PL) spectra (Figure 2). In the thin film of mithrene, the simulated and experimental reflectance spectra both show an absorption peak at the exciton wavelength confirming the unhybridized nature of system, and the PL is observed at the same wavelength demonstrating the direct band gap nature of mithrene. However, when the mithrene thickness is increased to 486 nm, new absorptive modes are observed both above and below the exciton which are the UEP and LEP, respectively. Sub-band gap emission was also observed in the thicker flakes in the PL because the LEP can emit due to its part-exciton characteristics. In addition to the UEP and LEP, the thicker mithrene flakes also hosted a higher-order (HO) mode similar to what we previously observed in perovskites of comparable thicknesses[45]. The emergence of the HO mode is because the mithrene is thick enough to support multiple orders of the lossy Fabry-Perot cavity. Therefore, there exists a higher order mode of the cavity at shorter wavelengths that is also coupling to the exciton. Since the HO mode-exciton coupling is strong enough for anticrossing to occur, and the two states have a large degree of detuning (energy different between the unperturbed exciton and cavity modes), the HO mode is slightly redshifted from the exciton peak. The HO mode is also mostly excitonic with a small amount of hybridization with light which is why it emits more than the LEP. The detuning between the HO mode and the exciton can be decreased by increasing the mithrene thickness (increasing the cavity wavelength) which increases the fraction of the cavity state in the HO mode. Eventually, the state would be approximately half-exciton, half-cavity making it a LEP. This continuous transition between HO mode can be clearly seen in simulations where the HO continuously redshifts with increased mithrene thickness until it is called a LEP (Supporting Information Figure S2).

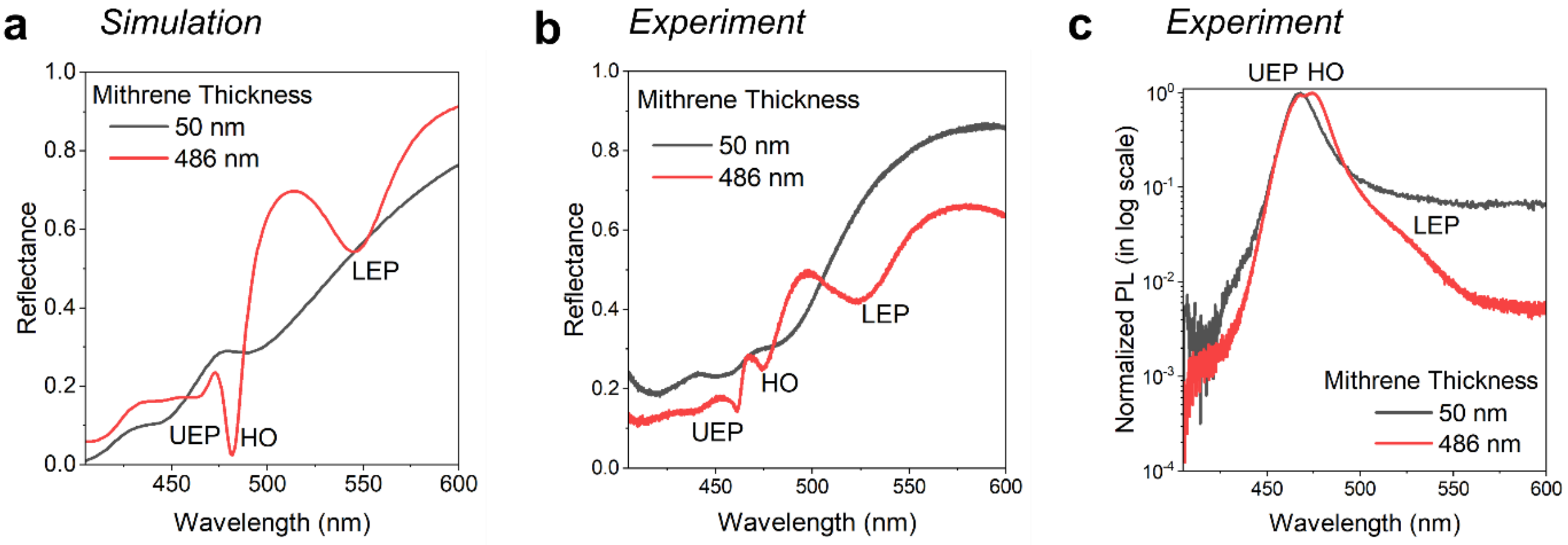

**Figure 2. Room-temperature exciton-polaritons in mithrene.** (a) Transfer-matrix calculation from the mithrene on the Au substrate showing the reflectance dips corresponding to the exciton and exciton-polariton (UEP, HO, and LEP) states emerging in 50 nm and 486 nm thick mithrene, respectively. (b) experimental observation of the exciton and exciton-polaritons in the reflectance spectrum matches closely with the simulation data. The shift in the polariton branches between experiment and simulation is attributed to the thickness-estimation error in the AFM measurement. (c) Photoluminescence spectroscopy showing exciton emission at 468 nm, while exciton-polariton shows emission at 468 nm (UEP), 474 nm (HO mode), and 527 nm (LEP).

The emissive properties of exciton-polaritons in mithrene are further investigated by comparing mithrene in open and closed cavity systems. From top to bottom, the open cavity system is mithrene (943 nm)/$Al_2O_3$ (10 nm)/Ag (100 nm) (Figure 3a), and the closed cavity is PMMA (~250 nm)/Ag (15 nm)/$Al_2O_3$ (10 nm)/mithrene (943 nm)/$Al_2O_3$ (10 nm)/Ag (100 nm) (Figure 3b). Ag is used as the metal in both systems since it is less absorptive than Au in the visible range (Supporting Information Figure S3), and the Ag substrate has 10 nm of $Al_2O_3$ on top that is deposited using atomic layer deposition (ALD) to prevent oxidation. For the closed cavity system, 10 nm of ALD $Al_2O_3$ is deposited on top of mithrene to protect it during the Ag sputtering process, and PMMA is spin-coated on top of the second Ag layer to prevent oxidation. By depositing Ag on top to make the closed cavity system, the Q-factor of the largest PL peak is increased by a factor of 2.26 compared to the open cavity system since the top interface of the lossy Fabry-Perot cavity is more reflective than the open cavity system (Supporting Information Figure S4). The increased Q-factor is seen as a narrowing of the peaks in both the reflectance (Figure 3c) and room temperature PL (Figure 3d). Additionally, the coupling parameter increases to 385 meV in the closed cavity system due to the decrease in cavity mode volume and cavity loss (Supporting Information Figure S5) which causes the multiple LEP modes in the closed cavity system to redshift from their open cavity wavelengths. The closed cavity system was also cooled down to 80 K, and its linewidths further decreases to sub-nanometer values due to decreased photon scattering (Figure 3e). The linewidth of an LEP is inversely related to its lifetime, and the lifetime of the LEP is a weighted average of the cavity and exciton lifetimes[51]. At room temperature, we hypothesize that the nonradiative lifetime of the exciton plays a significant role in the LEP lifetime. Therefore, as the temperature is reduced, the nonradiative lifetime of the exciton is prolonged, and the radiative lifetime of the cavity dominates the lifetime of the LEP. This is further confirmed by the PL at 80 K of mithrene on a quartz substrate. The transmissive property of the substrate prevents cavity modes from forming so its emission is purely excitonic. The exciton emission shows a larger linewidth than the LEP emission indicating that the high-order Fabry-Perot cavity modes reduce the linewidth of the LEP.

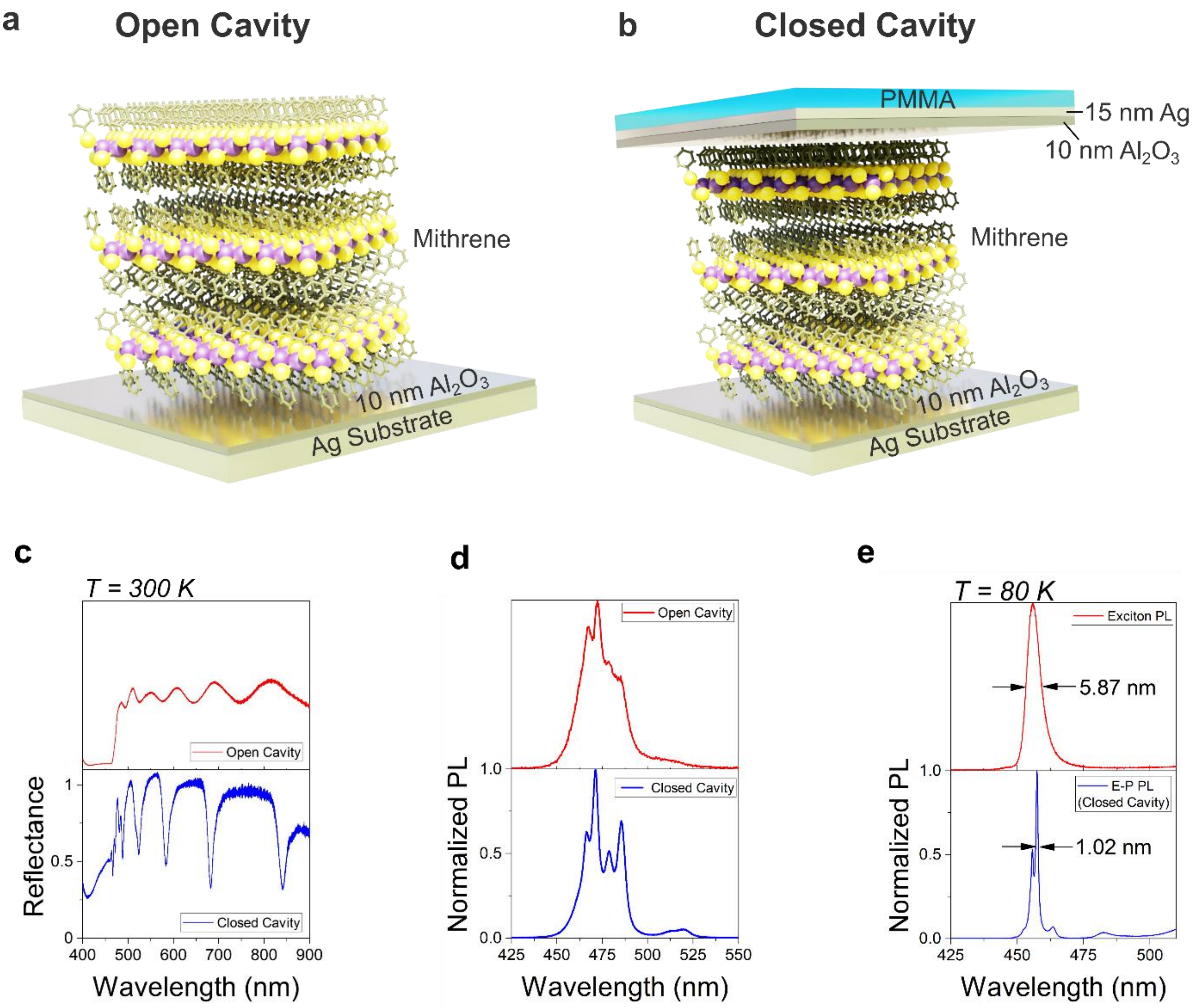

**Figure 3. Exciton-Polaritons in Open and Closed Cavity Mithrene.** Here, open cavity refers to mithrene/10 nm $Al_2O_3$/100 nm Ag (a) and closed cavity refers to PMMA/15 nm Ag/10 nm $Al_2O_3$/PMMA/ mithrene/10 nm $Al_2O_3$/100 nm Ag (b). Reflectance (c) and PL (d) from open and closed cavities. (e) The PL recorded at 80 K shows a linewidth narrowing of the E-P emission from closed cavity compared to the exciton emission.

Upon further cooling of the closed cavity system to < 50 K, the anti-crossing of exciton-polaritons is observed in the angle-dependent PL (Figure 4a, b)[52]. The emission at temperatures of 40 K and below is found to extend over a substantial portion of the visible range of light to 725 nm which is significantly longer than the emission at 80 K and room temperature where the longest wavelength emission is 520 nm. The states with the longest wavelength are found to be the most emissive, (excluding the unhybridized exciton emission), at 40 K and 20 K. Since the pump wavelength is above the band gap of mithrene, it creates a population of excitons that are then hybridized into exciton-polaritons. The excitons then relax into the LEP branches, but this process requires the presence of phonons. Therefore, this process occurs more quickly at 40 K than at 20 K which is why the longest wavelength emission at 40 K is more intense than the one at 20 K.

The lifetime of multiple LEP branches are also studied in both the open and closed cavity systems using time-resolved photoluminescence (TRPL) (Figure 4c and d). The lifetime in both systems are found to increase as the LEP redshifted and became less excitonic. This is because in both the open and closed cavity systems, the cavity mode has a longer lifetime than the exciton lifetime (0.27 ns). The closed cavity system shows an enhancement in lifetime over the open cavity by a factor of 1.86 (Supporting Information Figure S4). The biggest enhancement is seen at 454 nm compared to other emission wavelengths of the exciton-polaritons.

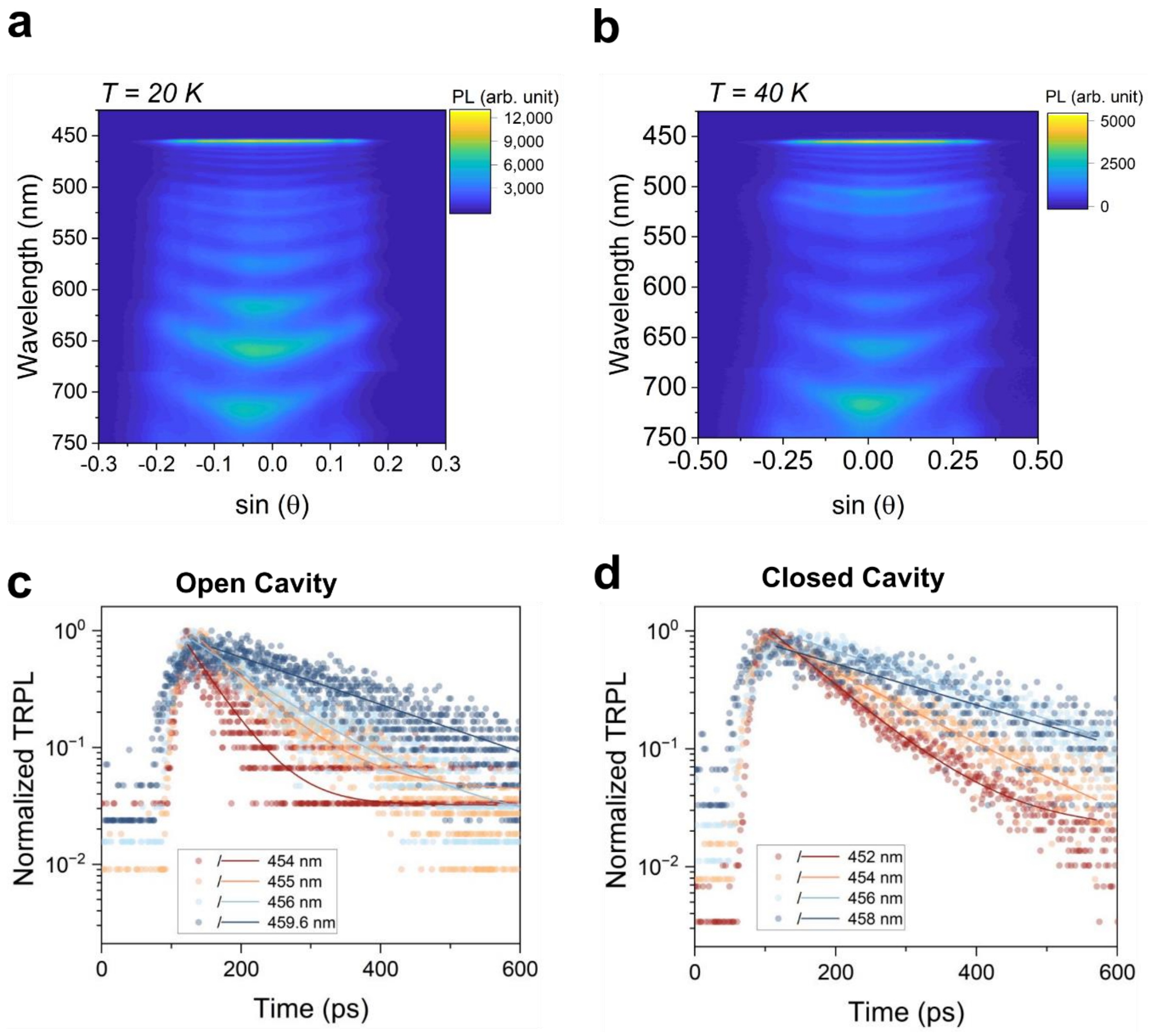

**Figure 4. Exciton-Polariton Dispersion and Lifetimes.** Temperature dependent E-k studies from mithrene in open cavity shows a clear dispersion from exciton-polaritons at temperatures of (a) 20 K and (b) 40 K. Time-resolved photoluminescence studies from (c) open and (d) closed cavity shows the exciton-polariton lifetime gets stretched by more than 2-fold in a closed cavity system.

## Conclusion

In conclusion, the strength of light-matter interactions in mithrene are studied in both open and closed cavity systems. Mithrene is found to host self-hybridized exciton-polaritons in the USC regime as $g/E_x = 0.14$ which is a record value for non-organic semiconductors in the visible spectrum. The formation of exciton-polariton states not only alters the optical dispersion of mithrene, but it also prolongs the lifetime of the states, enabling ≈1 nm PL linewidths. Additionally, at temperatures below 40 K, emissive exciton-polaritons from 452 nm to 752 nm are observed suggesting the potential for broadband, polaritonic LEDs and lasers. Our results show that mithrene, and perhaps even other MOCHA compounds, has excellent potential as an excitonic material for dispersion engineering in devices as well as fundamental studies of strongly coupled quantum photonic phenomena in the visible range.

## Methods

**Sample preparation.** Mithrene was prepared by biphasic interfacial synthesis as described previously[26]. Briefly, 3 mM solutions of silver nitrate in water and diphenyl diselenide are prepared separately, and then layered in a vial or vessel. Crystals evolve at the liquid-liquid interface and can be isolated by removing the water from the vessel by pipette, swirling the vial to cause the thin film to become adherent to the glass, and then decanting the remaining toluene. Small crystals grown from an aqueous layer yield typical final thicknesses of approximately 50 nm. Larger mithrene crystals were prepared by addition of ethylamine to the aqueous layer, and these crystals can be grown to hundreds of nanometers in thickness[53,54]. In a departure from Paritmongkol's method, the synthesis was performed at room temperature.

Mithrene is easily suspended and stable in a variety of organic alcohol solvents and can be drop cast onto desired substrates for imaging or optical characterization[26]. To understand the excitonic properties (absorption and PL) from the mithrene system, we drop-cast 50 µl of solvent with suspended mithrene crystals quartz substrates and allowed the solvent to evaporate at room temperature. For open cavity samples, the larger mithrene crystals from biphasic-ethylamine route were drop-cast on the 100-nm-thick Au substrate and the 10 nm $Al_2O_3$/100 nm Ag, separately. In the latter case, 10 nm $Al_2O_3$ inhibited any adverse reaction between the synthesized mithrene crystal and the Ag substrate. Both Ag and Au substrate were prepared by template-stripped process[55]. Further, 10 nm $Al_2O_3$/10 nm Ag followed by PMMA layer was deposited on the open cavity samples to realize a closed cavity system. Atomic layer deposition process was used to coat $Al_2O_3$ on the Ag substrate.

**Low-temperature PL and reflectance.** The reflectance and PL were recorded in reflection mode using Horiba LabRam HR Evolution confocal microscope. The white light intensity reflected by a polished silver mirror was used to normalize the data recorded from the mithrene sample for reflectance measurement. Photoluminescence spectra were recorded using a continuous wave excitation source at 405 nm and passing the emission through 600 grooves/mm before reaching the detector. For low-temperature measurements (up to 80 K), the samples were placed in a Linkam cryostat and cooled down to the desired temperature by controlling the flow of liquid nitrogen.

**TRPL measurement.** TRPL measurements were performed using an 80MHz, 140 fs Ti:sapphire laser. The 800nm fundamental emission was doubled via SHG to provide fs pulses at 400nm which was guided onto the mithrene samples. The corresponding emission was first collected and sent into an Acton SpectraPro SP-2750 grating spectrometer with a 300gpmm grating and dispersed onto a Teledyne PyLoN CCD array. To measure the lifetime, the dispersed signal was sent from the spectrometer to a Hamamatsu Universal Streak Camera which provided the time resolved information of the emission spectra.

**Optical simulations.** Optical simulations were performed using python scripts for transfer matrix method simulations as described in literature[35]. The complex refractive indices used in these simulations were measured using a J.A Woollam M2000 spectroscopic ellipsometer and fitting the experimental results to a series of Lorentz oscillators.

**E-K measurement.** To measure the E-K spectrum for mithrene, the sample was mounted in a Montana Instruments S200 cryostation with a 100X, 0.75 NA in-situ objective. The sample was excited with a 140fs, 80 MHz pulsed laser centered at 400nm. The angle-resolved emission was collected by imaging the back focal plane of the objective onto the entrance slit of a Princeton Instruments Isoplane SCT320 spectrometer using a 4f lens relay. The laser was filtered out by a 420nm long pass filter placed within the lens relay. The signal was then dispersed using a 300gpmm grating onto a PyLoN 400-BRX CCD array. Once collected, the data was processed by method described in the supplementary. The raw and processed data can be seen in figure S8.

**Acknowledgement.** D. J., S.B., J. L. and B.C. acknowledge partial support from Asian Office of Aerospace Research and Development of the Air Force Office of Scientific Research (AFOSR) (FA2386-21-1-4063), and the Office of Naval Research (N00014-23-1-203). J.N.H. and M. A. were supported by the US Department of Energy Integrated Computational and Data Infrastructure for Scientific Discovery grant DE-SC0022215. The research performed by C.E.S. at the Air Force Research Laboratory was supported by contract award FA807518D0015. J.R.H. acknowledges support from the Air Force Office of Scientific Research (Program Manager Dr. Gernot Pomrenke) under award number FA9550-20RYCOR059. A.M. and C.M. are supported by NSF-CAREER grant and NSF Intern program. T.P.D. and P.J.S. gratefully acknowledge support by Programmable Quantum Materials, an Energy Frontier Research Center funded by the US Department of Energy, Office of Science, Basic Energy Sciences, under award DE-SC0019443. The authors thank Christopher Chen for ellipsometry support. Work at the Molecular Foundry was supported by the Office of Science, Office of Basic Energy Sciences, of the U.S. Department of Energy under Contract No. DE-AC02-05CH11231.

Supporting Information

# Ultrastrong Light-Matter Coupling in 2D Metal-Organic Chalcogenolates

Surendra B. Anantharaman[1,*,†], Jason Lynch[1,†], Mariya Aleksich[2,3], Christopher E. Stevens[4,5], Christopher Munley[6], Bongjun Choi[1], Sridhar Shenoy[1], Thomas Darlington[7], Arka Majumdar[6,8], P. James Shuck[7], Joshua Hendrickson[5], J. Nathan Hohman[2,3], Deep Jariwala[1,*]

[1] Department of Electrical and Systems Engineering, University of Pennsylvania, Philadelphia, Pennsylvania 19104, United States

[2] Institute of Materials Science, University of Connecticut, Storrs CT 06269

[3] Department of Chemistry, University of Connecticut, Storrs CT 06269

[4] KBR Inc., Beavercreek, Ohio 45431, United States

[5] Air Force Research Laboratory, Sensors Directorate, Wright-Patterson Air Force Base, Ohio 45433, United States

[6] Department of Physics, University of Washington, Seattle, Washington 98195, United States

[7] Department of Mechanical Engineering, Columbia University, New York, new York 10027, United States

[8] Department of Electrical and Computer Engineering, University of Washington, Seattle, Washington 98195, United States

[*] Corresponding authors: sba@iitm.ac.in, dmj@seas.upenn.edu

[†] These authors contributed equally to this work.

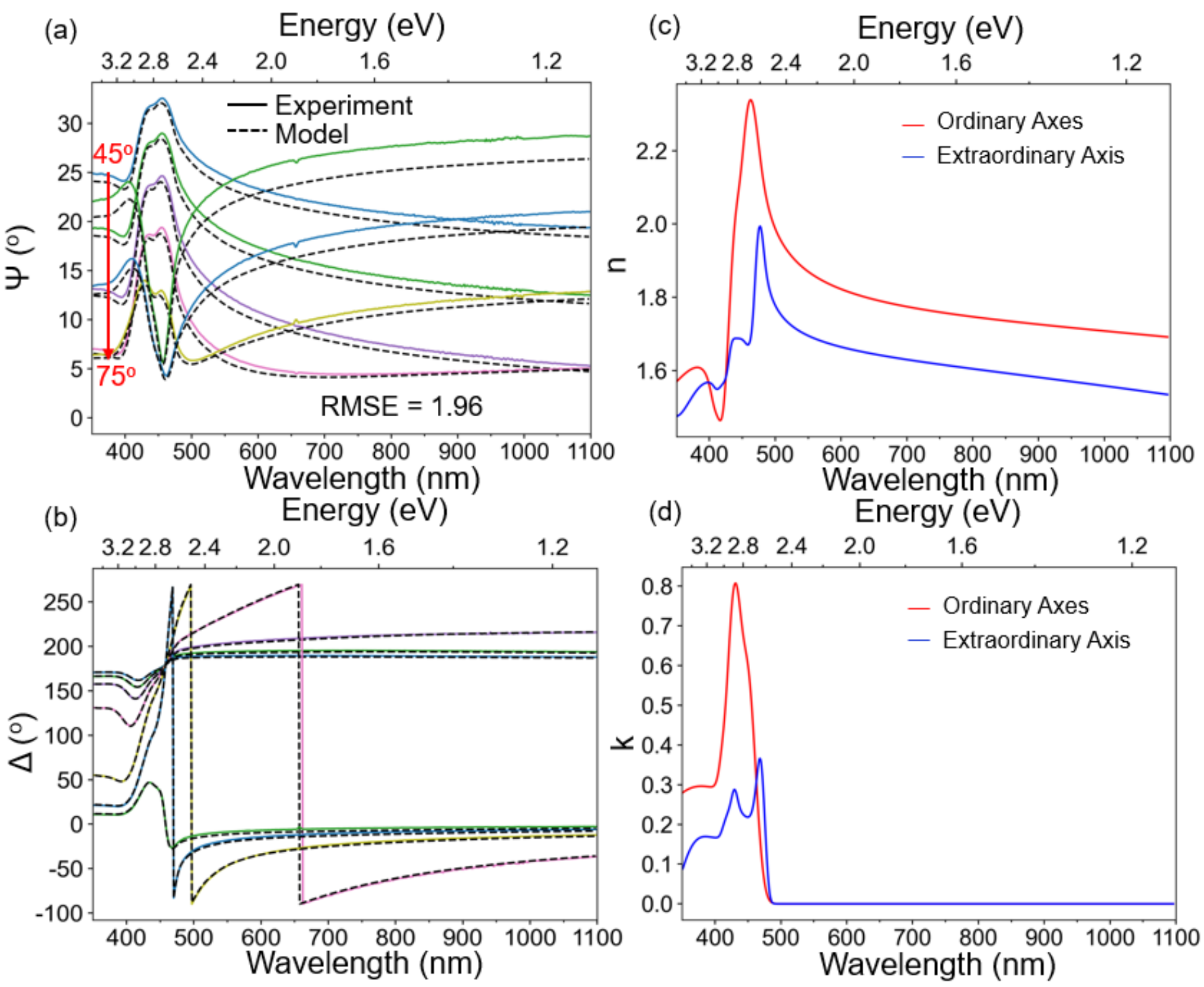

**Figure S1. Complex refractive index of mithrene** (a) Ψ and (b) Δ spectra that were measured using a J. A. Woollam M2000 spectroscopic ellipsometer and fitted to a multi-Lorentzian to extract the complex refractive index of mithrene. The model yielded a root-mean-square-error (RMSE = $\sqrt{\frac{1}{2p-q}\sum_i\left(\Psi_i^{Exp}-\Psi_i^{Model}\right)^2+\left(\Delta_i^{Exp}-\Delta_i^{Model}\right)^2}$ where p is the number of wavelengths measured, q is the number of fit parameters, and the sum is over all wavelengths) of 1.96 showing excellent agreement between experiment and model.

## Section 1. Weak-, Strong-, and Ultrastrong Light-Matter Coupling in Exciton-Polaritons

The Hamiltonians of exciton-polaritons can be described as the sum of the Hamiltonian of the unperturbed exciton ($H_{exciton} = \frac{1}{2}E_x\sigma_z$ where $E_x$ is the exciton energy and $\sigma_z$ is the exciton transition operator), the unperturbed cavity ($H_{cavity} = E_c a^\dagger a$ where $E_c$ is the cavity energy and $a^\dagger$ (a) is the creation (destruction) operator for a photon in the cavity), and the interaction Hamiltonian ($H_{int}$)[1]. In the cases of weak coupling (WC) and strong coupling (SC), only the first-order effects of an exciton absorbing and emitting a photon have to be considered for $H_{int}$. Therefore, the interaction Hamiltonian for WC and SC can be expressed as[1]:

$$H_{int} = g(\sigma_+ a + a^\dagger \sigma_-) \quad (1)$$

Where g is the coupling parameter and $\sigma_+$ ($\sigma_-$) is the absorption (emission) operator of the exciton. It is clear then that the coupling parameter can be interpretted as the rate at which energy oscillates between the exciton and cavity modes. When solving this Hamiltonian, called the Jaynes-Cummings Hamiltonian, under the condition of zero detuning ($\Delta = E_x - E_c$), the complex eigenenergies are found to be[2]:

$$E_{\pm} = E_x - i\frac{\Gamma_x + \Gamma_c}{4} \pm \sqrt{g^2 - \frac{(\Gamma_x - \Gamma_c)^2}{16}} \tag{2}$$

Where $\Gamma_x$ ($\Gamma_c$) is the loss rate of the exciton (cavity). Through inspection of Eq. 2, the WC regime is defined as the region where $g^2 < \frac{(\Gamma_x - \Gamma_c)^2}{16}$. In this regime, the coupling parameter only affects the imaginary part of the eigenenergy, and therefore, it only affects the decay rate of the states and does not affect their energies. However, when $g^2 > \frac{(\Gamma_x - \Gamma_c)^2}{16}$, the system enters the SC regime, and the states become hybrized as the coupling parameter shifts the energies of the two eigenstates.

Weakly and strongly coupled systems can both be accurately modelled while only considering the first order effects. However, as the coupling parameter increases, and energy oscillates between the exciton and cavity more rapidly, higher-order effects such as the fast-rotating components of the exciton-cavity interactions must be considered to accurately model the system. Although these effects are always present in exciton-polaritons, their effects are not significant until the coupling parameter is 10% of the exciton energy. Therefore, by convention, the ultrastrong coupling (USC) regime is when $\frac{g}{E_x} > 0.1$ . In the USC regime, the polariton energies are the positive solutions to the bi-quadratic[3]:

$$\left(E_{\pm}^2 - E_c^2\right)\left(E_{\pm}^2 - E_x^2\right) - \frac{4g^2 E_{\pm}^2 E_c}{E_x} = 0 \tag{3}$$

The coupling parameter for both the SC and USC regimes can then be extracted by using the transfer matrix method (TMM) to calculate the thickness dependence of the UEP and LEP since the cavity wavelength depends linearly on the mithrene thickness. The dispersion of the UEP and LEP are then fitted to Eq. 2 for the SC regime and Eq. 3 for the USC regime using a least-squared-error method to determine the coupling parameter of the exciton-polaritons.

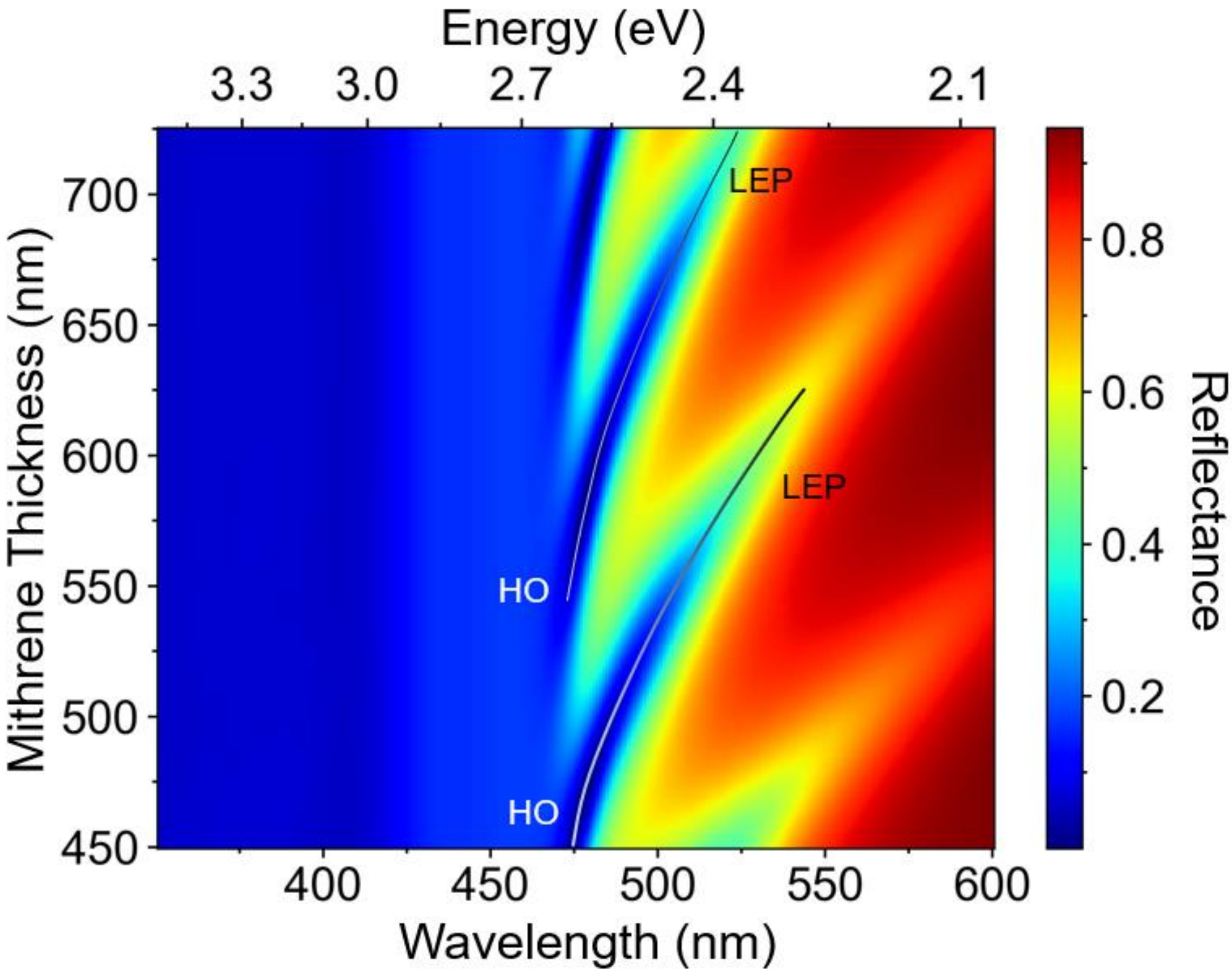

**Figure S2. Simulated higher order (HO) and lower exciton-polariton (LEP) modes in mithrene on Au.** Mithrene thickness-dependent reflectance spectra for a mithrene flake on an Au substrate. The HO mode can be seen to convert into an LEP mode as the mithrene thickness increases since the detuning between the exciton and cavity modes decreases.

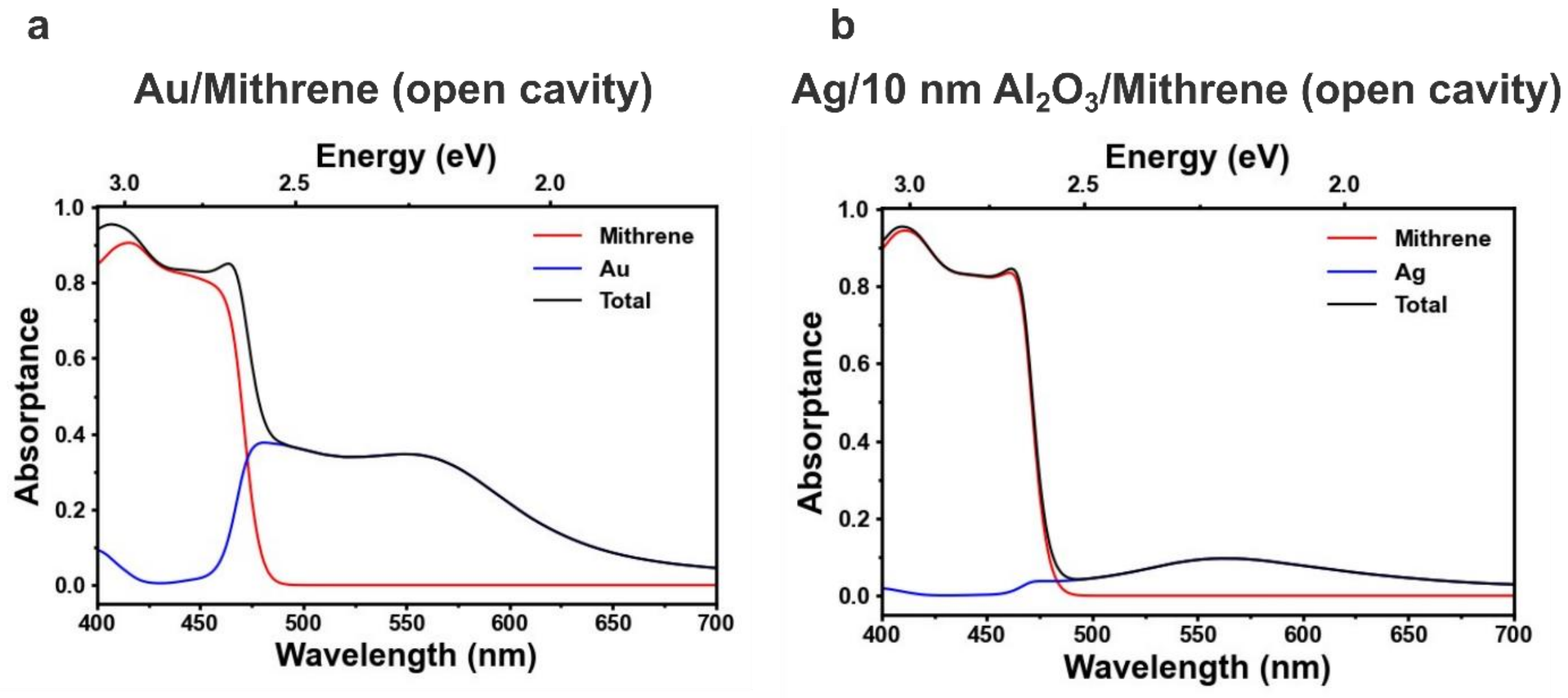

**Figure S3. Simulation studies on layer-dependent absorption for open-cavity polaritons.** (a) Using Au substrate for open cavity polaritons results in almost 40% light absorption around the mithrene exciton peak (475 nm). (b) By using Ag with a 10 nm $Al_2O_3$, the contribution from metal absorption can be drastically reduced from 40% to ~5%. Also, the 10 nm $Al_2O_3$ layer protects the silver from oxidation and avoids unwanted reaction between Ag and mithrene.

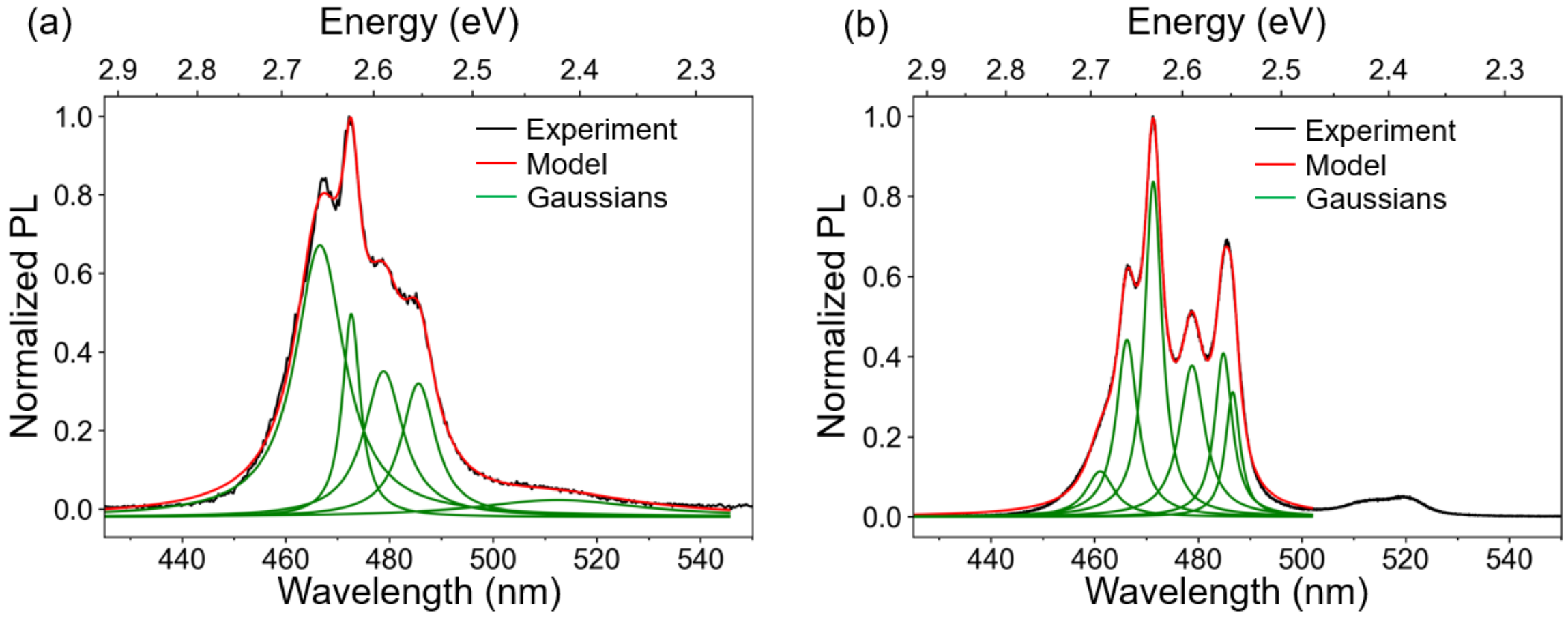

**Figure S4. Q-factor of exciton-polaritons in open and closed geometries.** The deconvoluted spectra of the (a) open and (b) closed cavity systems as shown in Figure 3d. The experimental data was deconvoluted into multiple Gaussian peaks and the average Q-factors were found to be 23.6 and 44.0 for the open and closed cavity systems, respectively.

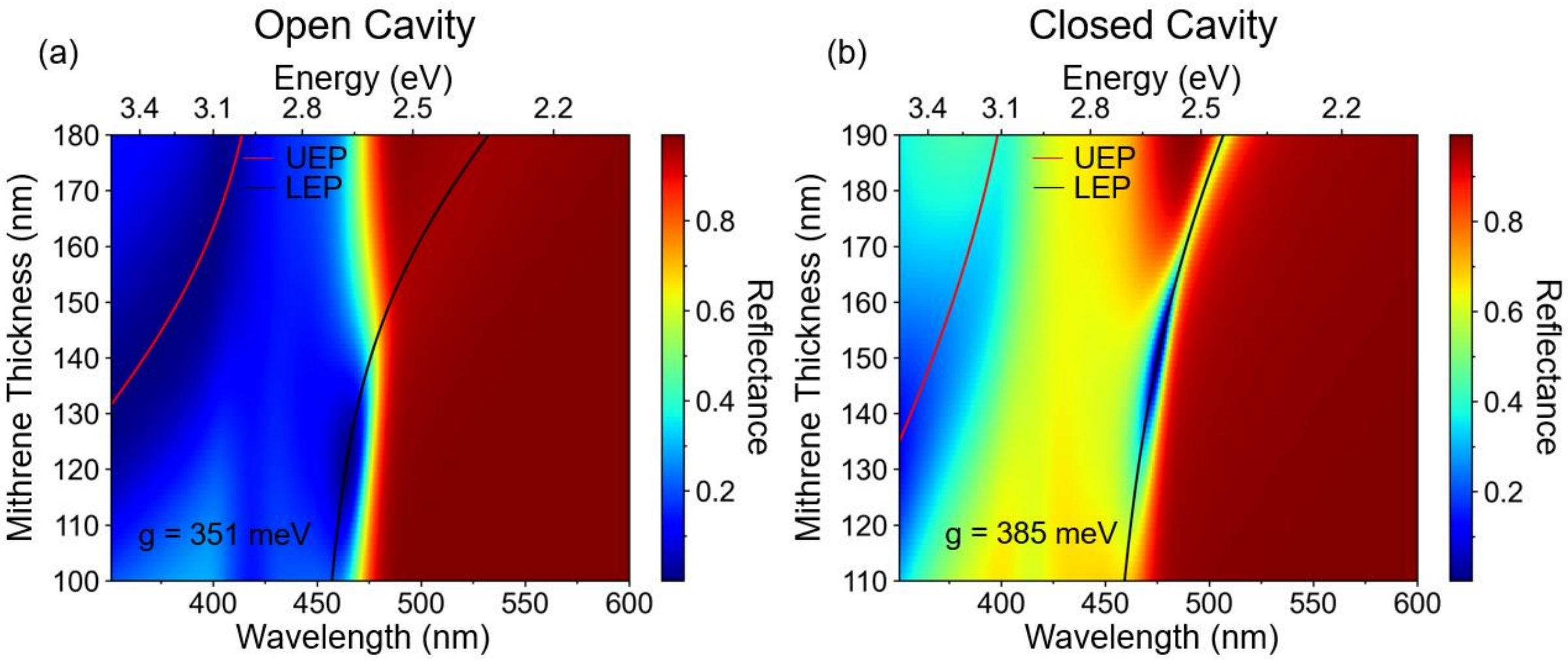

**Figure S5. Dispersion of exciton-polaritons in mithrene using silver reflectors.** The mithrene thickness dependence of the reflectance spectra for an (a) open and (b) closed cavity system. From top to botton, the open cavity system is mithrene/$Al_2O_3$ (10 nm)/Ag, and the closed cavity system is Ag (15 nm)/$Al_2O_3$ (10 nm)/mithrene/$Al_2O_3$ (10 nm)/Ag. The coupling parameters were extracted using the quantum Rabi model as discussed in section S1.

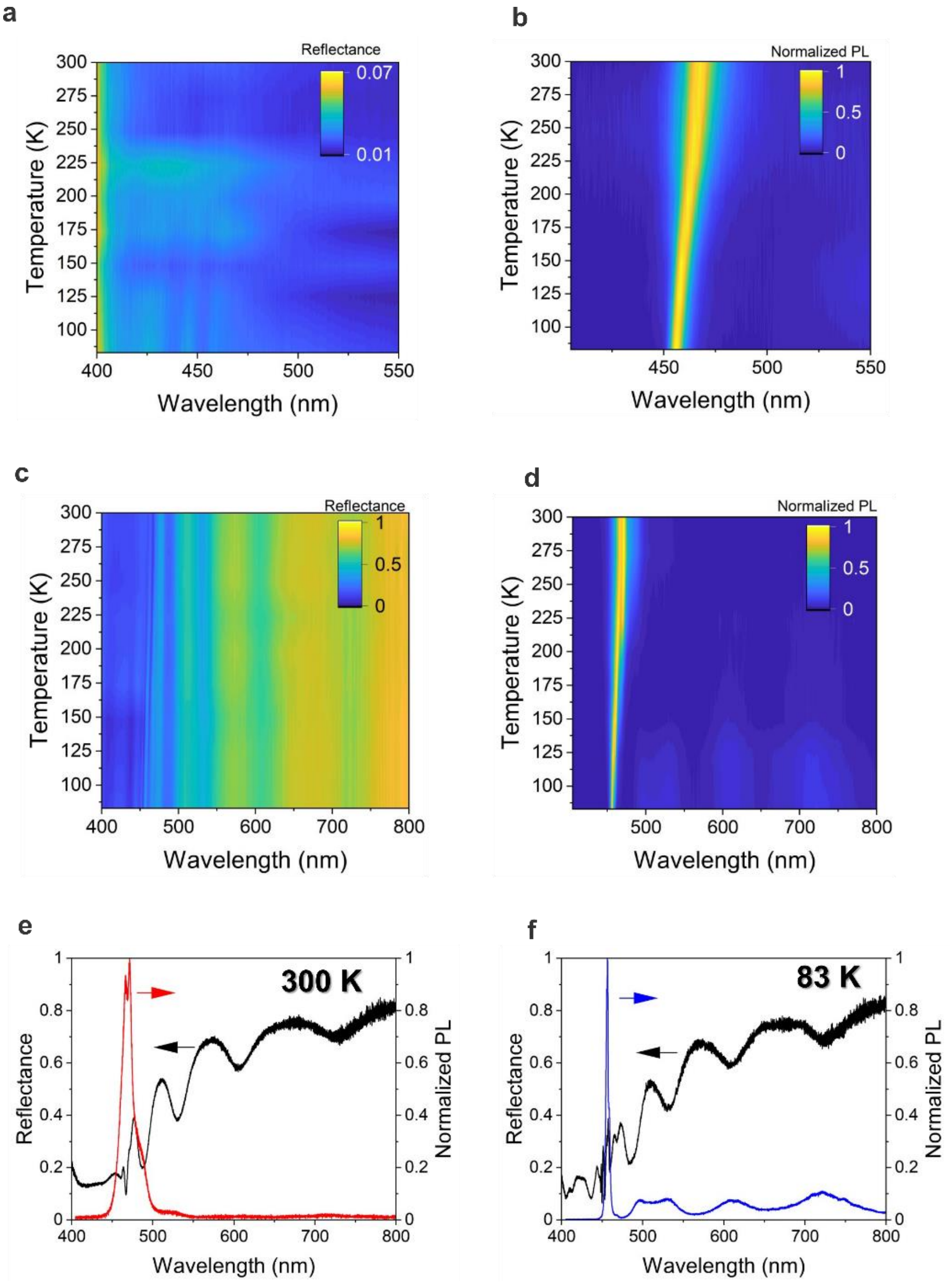

**Figure S6. Temperature-dependent optical properties of exciton and exciton-polariton from mithrene.** (a) reflectance and (b) PL spectra from the mithrene on quartz substrate shows a blue-shift in the exciton peak position upon cooling. (c) reflectance and (d) PL from the mithrene on Au substrate shows multiple polariton branches below the exciton peak due to the contribution from higher-order cavities. The line spectra at 300 K (e) and 80 K (f) shows the linewidth narrowing of the E-P peak and multiple polariton branches upon cooling to 80 K.

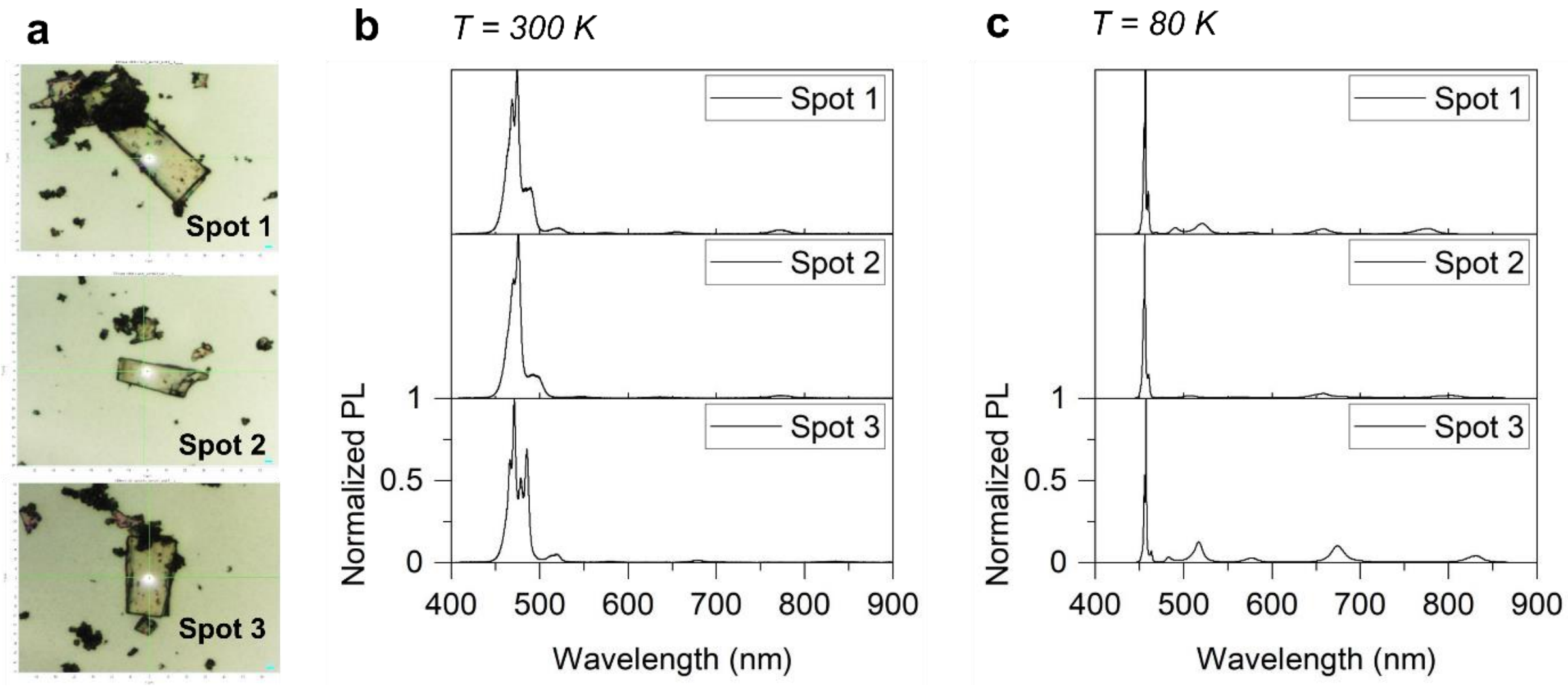

**Figure S7. Emergence of multiple polariton branches from Mithrene in closed cavity.** (a) Optical micrographs of mithrene in closed cavity (PMMA/15 nm Ag/10 nm $Al_2O_3$/PMMA/ Mithrene/10 nm $Al_2O_3$/100 nm Ag). (b) PL at 300 K showing a dominant exciton-polariton emission and (c) at 80 K the emergence of multiple polariton branches at longer wavelength can be observed. The linewidth of the most intense PL peak at 80 K were ~1 nm.

## Slope correction to E-K measurements.

To correct for the slope in the E-K measurements caused by the slight tiltof the CCD array, a linear transform was performed on the raw data. By analyzing the spectral tilt, a slope was determined from the spectra and could be used as a correction value. The slope was measured on several spectra and an average value of 0.0176 sin(θ)/pixel was determined. This correlates to the spectra being tilted one vertical pixel approximately every 29 horizontal pixels. Once this was determined, a linear transform was performed on the dataset correcting the data only in the background region, outside of the E-K measurement. Before and after correction data can be seen in Figure S8, Supporting Information.

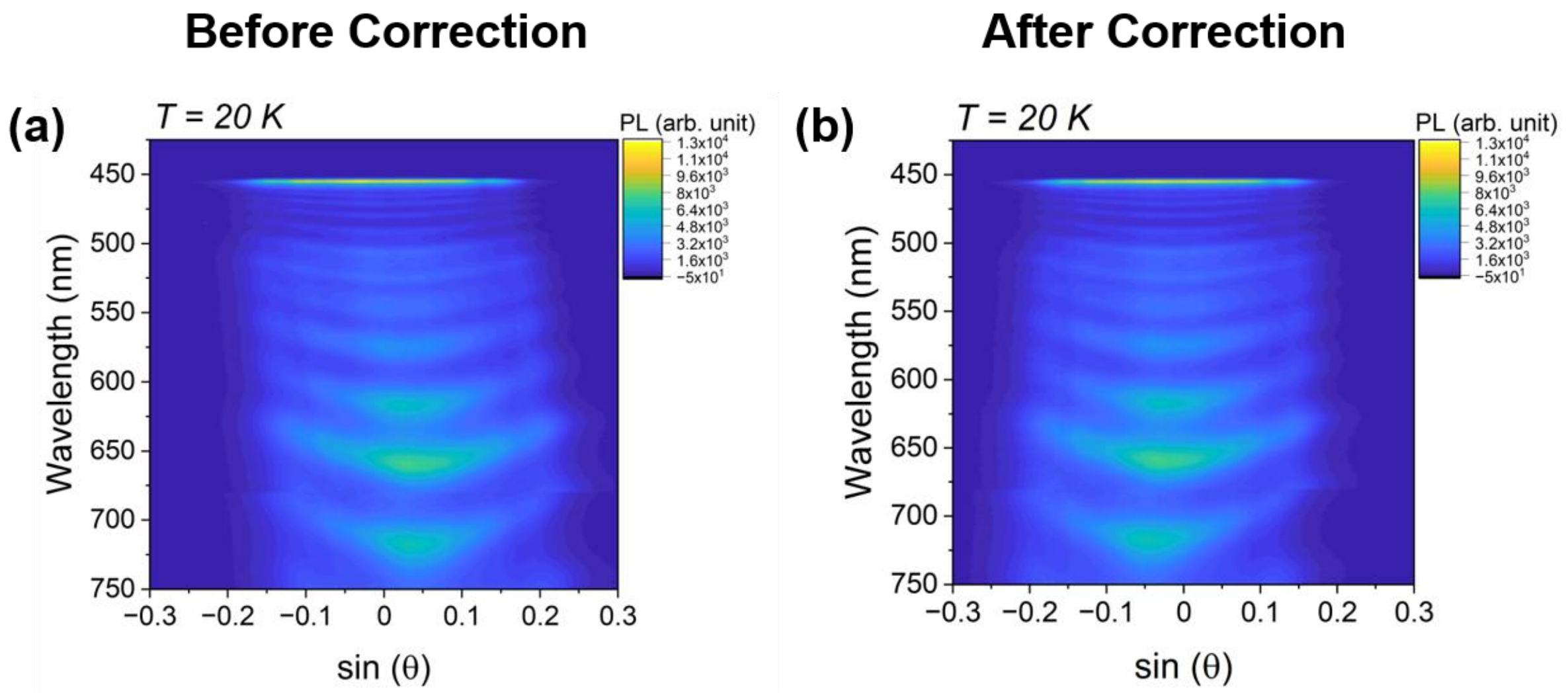

**Figure S8. E-k data correction.** The raw data without any data treatment (a) and correcting the drift by a slope value (b) is shown here.

**Table S1. Anisotropic, complex refractive index of mithrene**

| Wavelength (nm) | $n_{ord}$ | $k_{ord}$ | $n_{ext}$ | $k_{ext}$ |
|---|---|---|---|---|
| 350.9151 | 1.571173 | 0.280109 | 1.47571 | 0.088286 |
| 352.5063 | 1.574049 | 0.281995 | 1.476738 | 0.096357 |
| 354.0975 | 1.576889 | 0.283757 | 1.478456 | 0.104147 |
| 355.6887 | 1.579684 | 0.285395 | 1.480793 | 0.111581 |
| 357.2799 | 1.582424 | 0.28691 | 1.483671 | 0.1186 |
| 358.8712 | 1.5851 | 0.288304 | 1.48701 | 0.125164 |
| 360.4623 | 1.587702 | 0.289577 | 1.490733 | 0.131245 |
| 362.0535 | 1.590218 | 0.290732 | 1.494764 | 0.136829 |
| 363.6447 | 1.592637 | 0.291769 | 1.499033 | 0.141913 |
| 365.2359 | 1.594947 | 0.292691 | 1.503477 | 0.146499 |
| 366.8271 | 1.597135 | 0.2935 | 1.508036 | 0.150597 |
| 368.4183 | 1.599187 | 0.294198 | 1.512658 | 0.154222 |
| 370.0095 | 1.601087 | 0.294788 | 1.517293 | 0.157391 |
| 371.6007 | 1.602818 | 0.295273 | 1.521898 | 0.160125 |
| 373.1919 | 1.604362 | 0.295656 | 1.526432 | 0.162446 |
| 374.783 | 1.605699 | 0.29594 | 1.530859 | 0.164378 |
| 376.3742 | 1.606805 | 0.29613 | 1.535144 | 0.165947 |
| 377.9654 | 1.607654 | 0.29623 | 1.539258 | 0.167181 |
| 379.5565 | 1.608218 | 0.296245 | 1.543171 | 0.168108 |
| 381.1476 | 1.608461 | 0.296181 | 1.546858 | 0.168758 |
| 382.7388 | 1.608344 | 0.296046 | 1.550296 | 0.169163 |
| 384.3299 | 1.60782 | 0.295848 | 1.553463 | 0.169355 |
| 385.9211 | 1.606831 | 0.295597 | 1.556342 | 0.169369 |
| 387.5122 | 1.605308 | 0.295307 | 1.558918 | 0.169239 |
| 389.1033 | 1.60316 | 0.294997 | 1.561178 | 0.168997 |
| 390.6944 | 1.600273 | 0.294698 | 1.56311 | 0.168678 |
| 392.2855 | 1.596493 | 0.294465 | 1.564703 | 0.168313 |
| 393.8765 | 1.591627 | 0.294401 | 1.565946 | 0.16793 |
| 395.4676 | 1.585438 | 0.294691 | 1.566824 | 0.167553 |
| 397.0587 | 1.577678 | 0.295653 | 1.567308 | 0.167204 |
| 398.6497 | 1.568151 | 0.297783 | 1.567345 | 0.166901 |
| 400.2408 | 1.556831 | 0.301753 | 1.566839 | 0.166679 |
| 401.8318 | 1.543985 | 0.308339 | 1.565624 | 0.166635 |
| 403.4228 | 1.530261 | 0.318239 | 1.563501 | 0.167029 |
| 405.0138 | 1.516623 | 0.331817 | 1.560389 | 0.168361 |
| 406.6048 | 1.504095 | 0.348927 | 1.556575 | 0.17127 |
| 408.1957 | 1.493377 | 0.368962 | 1.552805 | 0.176184 |
| 409.7867 | 1.484573 | 0.391204 | 1.550012 | 0.182978 |
| 411.3776 | 1.477262 | 0.415348 | 1.548846 | 0.190997 |
| 412.9686 | 1.470976 | 0.441849 | 1.549418 | 0.199435 |
| 414.5595 | 1.465859 | 0.471819 | 1.551423 | 0.207695 |
| 416.1504 | 1.463119 | 0.506445 | 1.554388 | 0.21552 |
| 417.7413 | 1.46495 | 0.546182 | 1.557849 | 0.222964 |
| 419.3322 | 1.47391 | 0.590166 | 1.561461 | 0.230376 |
| 420.923 | 1.492035 | 0.636193 | 1.565181 | 0.238394 |
| 422.5139 | 1.520185 | 0.681265 | 1.569557 | 0.247783 |
| 424.1047 | 1.557893 | 0.722339 | 1.575944 | 0.258921 |
| 425.6955 | 1.603667 | 0.756908 | 1.586179 | 0.27101 |
| 427.2862 | 1.655419 | 0.783233 | 1.601455 | 0.281679 |
| 428.877 | 1.710808 | 0.800335 | 1.621014 | 0.287792 |
| 430.4678 | 1.767441 | 0.807924 | 1.641947 | 0.287228 |
| 432.0585 | 1.823002 | 0.806344 | 1.660556 | 0.280216 |
| 433.6492 | 1.875383 | 0.796555 | 1.674225 | 0.26914 |
| 435.2399 | 1.922865 | 0.780101 | 1.68241 | 0.257103 |
| 436.8305 | 1.964306 | 0.759002 | 1.686278 | 0.246446 |
| 438.4212 | 1.999361 | 0.735541 | 1.687603 | 0.238134 |
| 440.0118 | 2.028561 | 0.711942 | 1.687828 | 0.232058 |
| 441.6024 | 2.053274 | 0.689984 | 1.687746 | 0.227659 |
| 443.193 | 2.075454 | 0.670646 | 1.687621 | 0.224392 |
| 444.7835 | 2.097217 | 0.653906 | 1.687424 | 0.221898 |
| 446.3741 | 2.120349 | 0.638766 | 1.687012 | 0.220009 |
| 447.9646 | 2.145893 | 0.623543 | 1.686196 | 0.218721 |
| 449.5551 | 2.173941 | 0.606318 | 1.684783 | 0.218185 |
| 451.1456 | 2.203696 | 0.58541 | 1.682616 | 0.218731 |
| 452.736 | 2.233738 | 0.55971 | 1.679673 | 0.220888 |
| 454.3264 | 2.262382 | 0.528823 | 1.67619 | 0.225334 |
| 455.9168 | 2.287998 | 0.493026 | 1.672752 | 0.232775 |
| 457.5071 | 2.309237 | 0.45312 | 1.670309 | 0.243795 |
| 459.0975 | 2.325148 | 0.410229 | 1.670147 | 0.25876 |
| 460.6878 | 2.335189 | 0.365624 | 1.673921 | 0.277736 |
| 462.2781 | 2.339201 | 0.320592 | 1.683786 | 0.30026 |
| 463.8683 | 2.337339 | 0.276339 | 1.702363 | 0.32475 |
| 465.4586 | 2.330014 | 0.233935 | 1.732031 | 0.347754 |
| 467.0488 | 2.317833 | 0.194285 | 1.773331 | 0.363867 |
| 468.6389 | 2.301551 | 0.158107 | 1.823439 | 0.367236 |
| 470.2291 | 2.282022 | 0.12591 | 1.876279 | 0.354078 |
| 471.8192 | 2.260154 | 0.097992 | 1.924565 | 0.324384 |
| 473.4093 | 2.236852 | 0.074436 | 1.962292 | 0.281625 |
| 474.9993 | 2.212972 | 0.055119 | 1.986071 | 0.231067 |
| 476.5894 | 2.189265 | 0.039743 | 1.995016 | 0.178175 |
| 478.1794 | 2.166343 | 0.027878 | 1.99008 | 0.127858 |
| 479.7693 | 2.144657 | 0.019009 | 1.973672 | 0.084211 |
| 481.3592 | 2.124492 | 0.012592 | 1.949489 | 0.050097 |
| 482.9491 | 2.105987 | 0.008101 | 1.921966 | 0.026502 |
| 484.539 | 2.089153 | 0.005059 | 1.895155 | 0.012315 |
| 486.1288 | 2.073918 | 0.003067 | 1.87164 | 0.005002 |
| 487.7187 | 2.060149 | 0.001805 | 1.852246 | 0.001793 |
| 489.3084 | 2.047689 | 0.001031 | 1.836573 | 0.000595 |
| 490.8981 | 2.036373 | 0.000572 | 1.823773 | 0.000207 |
| 492.4879 | 2.026047 | 0.000308 | 1.813053 | 0.000091 |
| 494.0775 | 2.016572 | 0.000161 | 1.803841 | 0.000054 |
| 495.6671 | 2.007831 | 0.000082 | 1.79576 | 0.000038 |
| 497.2567 | 1.999722 | 0.000041 | 1.788564 | 0.000028 |
| 498.8463 | 1.992163 | 0.000019 | 1.782081 | 0.00002 |
| 500.4359 | 1.985087 | 0.000009 | 1.776189 | 0.000015 |
| 502.0253 | 1.978439 | 0.000004 | 1.770794 | 0.000011 |
| 503.6147 | 1.97217 | 0.000002 | 1.765825 | 0.000008 |
| 505.2042 | 1.966243 | 0.000001 | 1.761222 | 0.000006 |
| 506.7935 | 1.960625 | 0 | 1.756941 | 0.000004 |
| 508.3829 | 1.955288 | 0 | 1.752941 | 0.000003 |
| 509.9722 | 1.950207 | 0 | 1.749192 | 0.000002 |
| 511.5614 | 1.945361 | 0 | 1.745666 | 0.000002 |
| 513.1507 | 1.940731 | 0 | 1.74234 | 0.000001 |
| 514.7399 | 1.936302 | 0 | 1.739195 | 0.000001 |
| 516.329 | 1.932058 | 0 | 1.736212 | 0.000001 |
| 517.9181 | 1.927986 | 0 | 1.733379 | 0 |
| 519.5071 | 1.924075 | 0 | 1.730681 | 0 |
| 521.0961 | 1.920314 | 0 | 1.728108 | 0 |
| 522.6852 | 1.916693 | 0 | 1.725649 | 0 |
| 524.274 | 1.913204 | 0 | 1.723295 | 0 |
| 525.863 | 1.909838 | 0 | 1.721039 | 0 |
| 527.4518 | 1.906588 | 0 | 1.718873 | 0 |
| 529.0406 | 1.903448 | 0 | 1.716791 | 0 |
| 530.6294 | 1.900411 | 0 | 1.714786 | 0 |
| 532.2181 | 1.897472 | 0 | 1.712855 | 0 |
| 533.8068 | 1.894625 | 0 | 1.710992 | 0 |
| 535.3954 | 1.891866 | 0 | 1.709193 | 0 |
| 536.984 | 1.88919 | 0 | 1.707453 | 0 |
| 538.5725 | 1.886593 | 0 | 1.70577 | 0 |
| 540.161 | 1.884071 | 0 | 1.704139 | 0 |
| 541.7495 | 1.881621 | 0 | 1.702559 | 0 |
| 543.3378 | 1.879238 | 0 | 1.701025 | 0 |
| 544.9262 | 1.876921 | 0 | 1.699536 | 0 |
| 546.5145 | 1.874665 | 0 | 1.698089 | 0 |
| 548.1028 | 1.872469 | 0 | 1.696681 | 0 |
| 549.691 | 1.87033 | 0 | 1.695312 | 0 |
| 551.2792 | 1.868244 | 0 | 1.693978 | 0 |
| 552.8672 | 1.866211 | 0 | 1.692679 | 0 |
| 554.4553 | 1.864227 | 0 | 1.691412 | 0 |
| 556.0433 | 1.862291 | 0 | 1.690176 | 0 |
| 557.6313 | 1.860401 | 0 | 1.68897 | 0 |
| 559.2192 | 1.858554 | 0 | 1.687792 | 0 |
| 560.807 | 1.85675 | 0 | 1.686641 | 0 |
| 562.3949 | 1.854987 | 0 | 1.685516 | 0 |
| 563.9826 | 1.853263 | 0 | 1.684415 | 0 |
| 565.5703 | 1.851576 | 0 | 1.683338 | 0 |
| 567.158 | 1.849926 | 0 | 1.682283 | 0 |
| 568.7455 | 1.84831 | 0 | 1.68125 | 0 |
| 570.3331 | 1.846728 | 0 | 1.680238 | 0 |
| 571.9206 | 1.845179 | 0 | 1.679245 | 0 |
| 573.508 | 1.843661 | 0 | 1.678272 | 0 |
| 575.0954 | 1.842174 | 0 | 1.677316 | 0 |
| 576.6827 | 1.840715 | 0 | 1.676378 | 0 |
| 578.27 | 1.839285 | 0 | 1.675458 | 0 |
| 579.8572 | 1.837883 | 0 | 1.674553 | 0 |
| 581.4444 | 1.836506 | 0 | 1.673664 | 0 |
| 583.0315 | 1.835156 | 0 | 1.67279 | 0 |
| 584.6185 | 1.83383 | 0 | 1.671931 | 0 |
| 586.2055 | 1.832528 | 0 | 1.671085 | 0 |
| 587.7924 | 1.83125 | 0 | 1.670253 | 0 |
| 589.3793 | 1.829994 | 0 | 1.669435 | 0 |
| 590.9661 | 1.82876 | 0 | 1.66863 | 0 |
| 592.5529 | 1.827547 | 0 | 1.667836 | 0 |
| 594.1396 | 1.826355 | 0 | 1.667055 | 0 |
| 595.7262 | 1.825182 | 0 | 1.666286 | 0 |
| 597.3128 | 1.824029 | 0 | 1.665528 | 0 |
| 598.8993 | 1.822895 | 0 | 1.66478 | 0 |
| 600.4858 | 1.821779 | 0 | 1.664043 | 0 |
| 602.0721 | 1.820682 | 0 | 1.663316 | 0 |
| 603.6585 | 1.819601 | 0 | 1.6626 | 0 |
| 605.2448 | 1.818537 | 0 | 1.661892 | 0 |
| 606.8309 | 1.817489 | 0 | 1.661193 | 0 |
| 608.4171 | 1.816458 | 0 | 1.660504 | 0 |
| 610.0032 | 1.815442 | 0 | 1.659822 | 0 |
| 611.5892 | 1.81444 | 0 | 1.65915 | 0 |
| 613.1752 | 1.813454 | 0 | 1.658485 | 0 |
| 614.761 | 1.812482 | 0 | 1.657827 | 0 |
| 616.3469 | 1.811524 | 0 | 1.657178 | 0 |
| 617.9327 | 1.810579 | 0 | 1.656535 | 0 |
| 619.5184 | 1.809648 | 0 | 1.6559 | 0 |
| 621.1039 | 1.808729 | 0 | 1.655272 | 0 |
| 622.6896 | 1.807823 | 0 | 1.65465 | 0 |
| 624.275 | 1.806929 | 0 | 1.654035 | 0 |
| 625.8605 | 1.806048 | 0 | 1.653427 | 0 |
| 627.4458 | 1.805178 | 0 | 1.652824 | 0 |
| 629.0311 | 1.804319 | 0 | 1.652228 | 0 |
| 630.6163 | 1.803471 | 0 | 1.651637 | 0 |
| 632.2015 | 1.802635 | 0 | 1.651052 | 0 |
| 633.7866 | 1.801808 | 0 | 1.650473 | 0 |
| 635.3716 | 1.800993 | 0 | 1.649899 | 0 |
| 636.9565 | 1.800187 | 0 | 1.64933 | 0 |
| 638.5414 | 1.799391 | 0 | 1.648766 | 0 |
| 640.1262 | 1.798605 | 0 | 1.648208 | 0 |
| 641.711 | 1.797828 | 0 | 1.647654 | 0 |
| 643.2956 | 1.797061 | 0 | 1.647105 | 0 |
| 644.8802 | 1.796303 | 0 | 1.64656 | 0 |
| 646.4647 | 1.795554 | 0 | 1.64602 | 0 |
| 648.0492 | 1.794813 | 0 | 1.645484 | 0 |
| 649.6335 | 1.794081 | 0 | 1.644953 | 0 |
| 651.2179 | 1.793357 | 0 | 1.644425 | 0 |
| 652.8021 | 1.792641 | 0 | 1.643902 | 0 |
| 654.3862 | 1.791933 | 0 | 1.643383 | 0 |
| 655.9703 | 1.791232 | 0 | 1.642867 | 0 |
| 657.5543 | 1.79054 | 0 | 1.642356 | 0 |
| 659.1383 | 1.789854 | 0 | 1.641848 | 0 |
| 660.7221 | 1.789176 | 0 | 1.641343 | 0 |
| 662.3059 | 1.788505 | 0 | 1.640842 | 0 |
| 663.8896 | 1.787841 | 0 | 1.640345 | 0 |
| 665.4733 | 1.787183 | 0 | 1.639851 | 0 |
| 667.0568 | 1.786533 | 0 | 1.63936 | 0 |
| 668.6403 | 1.785889 | 0 | 1.638872 | 0 |
| 670.2236 | 1.785251 | 0 | 1.638387 | 0 |
| 671.8069 | 1.784619 | 0 | 1.637905 | 0 |
| 673.3902 | 1.783994 | 0 | 1.637427 | 0 |
| 674.9733 | 1.783375 | 0 | 1.636951 | 0 |
| 676.5564 | 1.782761 | 0 | 1.636478 | 0 |
| 678.1394 | 1.782153 | 0 | 1.636008 | 0 |
| 679.7223 | 1.781551 | 0 | 1.63554 | 0 |
| 681.3052 | 1.780955 | 0 | 1.635075 | 0 |
| 682.8879 | 1.780364 | 0 | 1.634613 | 0 |
| 684.4706 | 1.779778 | 0 | 1.634153 | 0 |
| 686.0532 | 1.779197 | 0 | 1.633695 | 0 |
| 687.6357 | 1.778622 | 0 | 1.63324 | 0 |
| 689.2181 | 1.778052 | 0 | 1.632788 | 0 |
| 690.8004 | 1.777486 | 0 | 1.632337 | 0 |
| 692.3827 | 1.776926 | 0 | 1.631889 | 0 |
| 693.9649 | 1.77637 | 0 | 1.631443 | 0 |
| 695.547 | 1.775819 | 0 | 1.630999 | 0 |
| 697.129 | 1.775273 | 0 | 1.630557 | 0 |
| 698.7109 | 1.774731 | 0 | 1.630118 | 0 |
| 700.2928 | 1.774193 | 0 | 1.62968 | 0 |
| 701.8745 | 1.77366 | 0 | 1.629244 | 0 |
| 703.4562 | 1.773131 | 0 | 1.62881 | 0 |
| 705.0377 | 1.772606 | 0 | 1.628378 | 0 |
| 706.6193 | 1.772086 | 0 | 1.627948 | 0 |
| 708.2006 | 1.771569 | 0 | 1.62752 | 0 |
| 709.782 | 1.771057 | 0 | 1.627093 | 0 |
| 711.3632 | 1.770548 | 0 | 1.626668 | 0 |
| 712.9443 | 1.770043 | 0 | 1.626245 | 0 |
| 714.5254 | 1.769542 | 0 | 1.625823 | 0 |
| 716.1063 | 1.769045 | 0 | 1.625403 | 0 |
| 717.6872 | 1.768551 | 0 | 1.624985 | 0 |
| 719.268 | 1.768061 | 0 | 1.624568 | 0 |
| 720.8487 | 1.767575 | 0 | 1.624152 | 0 |
| 722.4293 | 1.767092 | 0 | 1.623738 | 0 |
| 724.0098 | 1.766612 | 0 | 1.623326 | 0 |
| 725.5902 | 1.766136 | 0 | 1.622915 | 0 |
| 727.1705 | 1.765663 | 0 | 1.622505 | 0 |
| 728.7508 | 1.765193 | 0 | 1.622096 | 0 |
| 730.3309 | 1.764726 | 0 | 1.621689 | 0 |
| 731.911 | 1.764263 | 0 | 1.621283 | 0 |
| 733.4909 | 1.763802 | 0 | 1.620879 | 0 |
| 735.0708 | 1.763345 | 0 | 1.620475 | 0 |
| 736.6505 | 1.762891 | 0 | 1.620073 | 0 |
| 738.2302 | 1.762439 | 0 | 1.619672 | 0 |
| 739.8098 | 1.761991 | 0 | 1.619272 | 0 |
| 741.3893 | 1.761545 | 0 | 1.618873 | 0 |
| 742.9686 | 1.761102 | 0 | 1.618476 | 0 |
| 744.548 | 1.760662 | 0 | 1.618079 | 0 |
| 746.1271 | 1.760225 | 0 | 1.617684 | 0 |
| 747.7062 | 1.75979 | 0 | 1.617289 | 0 |
| 749.2853 | 1.759358 | 0 | 1.616895 | 0 |
| 750.8641 | 1.758929 | 0 | 1.616503 | 0 |
| 752.443 | 1.758502 | 0 | 1.616111 | 0 |
| 754.0216 | 1.758078 | 0 | 1.615721 | 0 |
| 755.6002 | 1.757656 | 0 | 1.615331 | 0 |
| 757.1788 | 1.757237 | 0 | 1.614942 | 0 |
| 758.7572 | 1.75682 | 0 | 1.614554 | 0 |
| 760.3354 | 1.756406 | 0 | 1.614167 | 0 |
| 761.9137 | 1.755994 | 0 | 1.61378 | 0 |
| 763.4918 | 1.755584 | 0 | 1.613395 | 0 |
| 765.0698 | 1.755177 | 0 | 1.61301 | 0 |
| 766.6477 | 1.754772 | 0 | 1.612626 | 0 |
| 768.2255 | 1.754369 | 0 | 1.612243 | 0 |
| 769.8032 | 1.753968 | 0 | 1.61186 | 0 |
| 771.3807 | 1.75357 | 0 | 1.611479 | 0 |
| 772.9583 | 1.753174 | 0 | 1.611098 | 0 |
| 774.5356 | 1.752779 | 0 | 1.610717 | 0 |
| 776.1129 | 1.752387 | 0 | 1.610337 | 0 |
| 777.6901 | 1.751996 | 0 | 1.609959 | 0 |
| 779.2672 | 1.751608 | 0 | 1.60958 | 0 |
| 780.8441 | 1.751221 | 0 | 1.609202 | 0 |
| 782.421 | 1.750836 | 0 | 1.608825 | 0 |
| 783.9978 | 1.750453 | 0 | 1.608449 | 0 |
| 785.5744 | 1.750072 | 0 | 1.608073 | 0 |
| 787.151 | 1.749692 | 0 | 1.607697 | 0 |
| 788.7274 | 1.749315 | 0 | 1.607322 | 0 |
| 790.3038 | 1.748939 | 0 | 1.606948 | 0 |
| 791.88 | 1.748565 | 0 | 1.606574 | 0 |
| 793.4561 | 1.748192 | 0 | 1.6062 | 0 |
| 795.0321 | 1.747822 | 0 | 1.605828 | 0 |
| 796.608 | 1.747453 | 0 | 1.605455 | 0 |
| 798.1838 | 1.747086 | 0 | 1.605083 | 0 |
| 799.7595 | 1.74672 | 0 | 1.604712 | 0 |
| 801.335 | 1.746356 | 0 | 1.604341 | 0 |
| 802.9105 | 1.745994 | 0 | 1.60397 | 0 |
| 804.4858 | 1.745633 | 0 | 1.6036 | 0 |
| 806.061 | 1.745274 | 0 | 1.60323 | 0 |
| 807.6362 | 1.744916 | 0 | 1.602861 | 0 |
| 809.2112 | 1.74456 | 0 | 1.602492 | 0 |
| 810.7861 | 1.744205 | 0 | 1.602123 | 0 |
| 812.3609 | 1.743852 | 0 | 1.601755 | 0 |
| 813.9355 | 1.7435 | 0 | 1.601387 | 0 |
| 815.5101 | 1.743149 | 0 | 1.601019 | 0 |
| 817.0846 | 1.7428 | 0 | 1.600652 | 0 |
| 818.6589 | 1.742453 | 0 | 1.600285 | 0 |
| 820.2331 | 1.742107 | 0 | 1.599918 | 0 |
| 821.8073 | 1.741762 | 0 | 1.599552 | 0 |
| 823.3812 | 1.741418 | 0 | 1.599186 | 0 |
| 824.9551 | 1.741076 | 0 | 1.59882 | 0 |
| 826.5288 | 1.740735 | 0 | 1.598455 | 0 |
| 828.1024 | 1.740395 | 0 | 1.598089 | 0 |
| 829.676 | 1.740057 | 0 | 1.597724 | 0 |
| 831.2493 | 1.73972 | 0 | 1.59736 | 0 |
| 832.8226 | 1.739384 | 0 | 1.596995 | 0 |
| 834.3958 | 1.739049 | 0 | 1.596631 | 0 |
| 835.9689 | 1.738716 | 0 | 1.596267 | 0 |
| 837.5418 | 1.738383 | 0 | 1.595903 | 0 |
| 839.1146 | 1.738052 | 0 | 1.595539 | 0 |
| 840.6873 | 1.737722 | 0 | 1.595176 | 0 |
| 842.2599 | 1.737393 | 0 | 1.594812 | 0 |
| 843.8323 | 1.737065 | 0 | 1.594449 | 0 |
| 845.4046 | 1.736738 | 0 | 1.594086 | 0 |
| 846.9769 | 1.736412 | 0 | 1.593723 | 0 |
| 848.549 | 1.736088 | 0 | 1.59336 | 0 |
| 850.1209 | 1.735764 | 0 | 1.592998 | 0 |
| 851.6927 | 1.735441 | 0 | 1.592635 | 0 |
| 853.2645 | 1.73512 | 0 | 1.592273 | 0 |
| 854.8361 | 1.734799 | 0 | 1.591911 | 0 |
| 856.4075 | 1.73448 | 0 | 1.591548 | 0 |
| 857.9788 | 1.734161 | 0 | 1.591186 | 0 |
| 859.55 | 1.733843 | 0 | 1.590824 | 0 |
| 861.1211 | 1.733526 | 0 | 1.590463 | 0 |
| 862.6921 | 1.733211 | 0 | 1.590101 | 0 |
| 864.263 | 1.732895 | 0 | 1.589739 | 0 |
| 865.8337 | 1.732581 | 0 | 1.589378 | 0 |
| 867.4043 | 1.732268 | 0 | 1.589016 | 0 |
| 868.9747 | 1.731956 | 0 | 1.588655 | 0 |
| 870.545 | 1.731644 | 0 | 1.588293 | 0 |
| 872.1152 | 1.731334 | 0 | 1.587932 | 0 |
| 873.6854 | 1.731024 | 0 | 1.58757 | 0 |
| 875.2552 | 1.730715 | 0 | 1.587209 | 0 |
| 876.8251 | 1.730407 | 0 | 1.586848 | 0 |
| 878.3948 | 1.730099 | 0 | 1.586486 | 0 |
| 879.9644 | 1.729793 | 0 | 1.586125 | 0 |
| 881.5338 | 1.729487 | 0 | 1.585764 | 0 |
| 883.1031 | 1.729182 | 0 | 1.585402 | 0 |
| 884.6724 | 1.728877 | 0 | 1.585041 | 0 |
| 886.2414 | 1.728574 | 0 | 1.58468 | 0 |
| 887.8102 | 1.728271 | 0 | 1.584318 | 0 |
| 889.3791 | 1.727969 | 0 | 1.583957 | 0 |
| 890.9478 | 1.727667 | 0 | 1.583596 | 0 |
| 892.5162 | 1.727367 | 0 | 1.583234 | 0 |
| 894.0846 | 1.727067 | 0 | 1.582873 | 0 |
| 895.6528 | 1.726767 | 0 | 1.582511 | 0 |
| 897.221 | 1.726468 | 0 | 1.58215 | 0 |
| 898.7889 | 1.72617 | 0 | 1.581788 | 0 |
| 900.3568 | 1.725873 | 0 | 1.581427 | 0 |
| 901.9245 | 1.725576 | 0 | 1.581065 | 0 |
| 903.4921 | 1.72528 | 0 | 1.580703 | 0 |
| 905.0596 | 1.724985 | 0 | 1.580341 | 0 |
| 906.6268 | 1.72469 | 0 | 1.579979 | 0 |
| 908.194 | 1.724396 | 0 | 1.579617 | 0 |
| 909.761 | 1.724102 | 0 | 1.579255 | 0 |
| 911.3279 | 1.723809 | 0 | 1.578893 | 0 |
| 912.8947 | 1.723516 | 0 | 1.578531 | 0 |
| 914.4613 | 1.723224 | 0 | 1.578168 | 0 |
| 916.0278 | 1.722933 | 0 | 1.577806 | 0 |
| 917.5941 | 1.722642 | 0 | 1.577443 | 0 |
| 919.1603 | 1.722352 | 0 | 1.57708 | 0 |
| 920.7264 | 1.722062 | 0 | 1.576717 | 0 |
| 922.2924 | 1.721773 | 0 | 1.576355 | 0 |
| 923.8582 | 1.721485 | 0 | 1.575991 | 0 |
| 925.4238 | 1.721196 | 0 | 1.575628 | 0 |
| 926.9894 | 1.720909 | 0 | 1.575265 | 0 |
| 928.5547 | 1.720622 | 0 | 1.574901 | 0 |
| 930.1199 | 1.720335 | 0 | 1.574538 | 0 |
| 931.6849 | 1.720049 | 0 | 1.574174 | 0 |
| 933.2499 | 1.719763 | 0 | 1.57381 | 0 |
| 934.8148 | 1.719478 | 0 | 1.573446 | 0 |
| 936.3794 | 1.719194 | 0 | 1.573082 | 0 |
| 937.9438 | 1.718909 | 0 | 1.572717 | 0 |
| 939.5082 | 1.718626 | 0 | 1.572353 | 0 |
| 941.0724 | 1.718342 | 0 | 1.571988 | 0 |
| 942.6365 | 1.718059 | 0 | 1.571623 | 0 |
| 944.2005 | 1.717777 | 0 | 1.571258 | 0 |
| 945.7643 | 1.717495 | 0 | 1.570893 | 0 |
| 947.3278 | 1.717214 | 0 | 1.570527 | 0 |
| 948.8914 | 1.716932 | 0 | 1.570162 | 0 |
| 950.4547 | 1.716652 | 0 | 1.569796 | 0 |
| 952.0179 | 1.716371 | 0 | 1.56943 | 0 |
| 953.5809 | 1.716091 | 0 | 1.569064 | 0 |
| 955.1439 | 1.715812 | 0 | 1.568698 | 0 |
| 956.7067 | 1.715533 | 0 | 1.568331 | 0 |
| 958.2692 | 1.715254 | 0 | 1.567964 | 0 |
| 959.8317 | 1.714975 | 0 | 1.567597 | 0 |
| 961.3939 | 1.714697 | 0 | 1.56723 | 0 |
| 962.9562 | 1.714419 | 0 | 1.566863 | 0 |
| 964.5182 | 1.714142 | 0 | 1.566495 | 0 |
| 966.08 | 1.713865 | 0 | 1.566127 | 0 |
| 967.6417 | 1.713589 | 0 | 1.565759 | 0 |
| 969.2032 | 1.713312 | 0 | 1.565391 | 0 |
| 970.7646 | 1.713036 | 0 | 1.565022 | 0 |
| 972.3259 | 1.712761 | 0 | 1.564654 | 0 |
| 973.887 | 1.712485 | 0 | 1.564284 | 0 |
| 975.4479 | 1.71221 | 0 | 1.563915 | 0 |
| 977.0087 | 1.711936 | 0 | 1.563546 | 0 |
| 978.5693 | 1.711661 | 0 | 1.563176 | 0 |
| 980.1299 | 1.711387 | 0 | 1.562806 | 0 |
| 981.6901 | 1.711114 | 0 | 1.562436 | 0 |
| 983.2504 | 1.71084 | 0 | 1.562066 | 0 |
| 984.8104 | 1.710567 | 0 | 1.561695 | 0 |
| 986.3702 | 1.710294 | 0 | 1.561324 | 0 |
| 987.93 | 1.710021 | 0 | 1.560953 | 0 |
| 989.4894 | 1.709749 | 0 | 1.560581 | 0 |
| 991.049 | 1.709477 | 0 | 1.56021 | 0 |
| 992.6081 | 1.709205 | 0 | 1.559838 | 0 |
| 994.1672 | 1.708934 | 0 | 1.559465 | 0 |
| 995.7262 | 1.708663 | 0 | 1.559093 | 0 |
| 997.2849 | 1.708392 | 0 | 1.55872 | 0 |
| 998.8434 | 1.708121 | 0 | 1.558347 | 0 |
| 1012.442 | 1.705767 | 0 | 1.555078 | 0 |
| 1015.843 | 1.705181 | 0 | 1.554257 | 0 |
| 1019.245 | 1.704595 | 0 | 1.553434 | 0 |
| 1022.646 | 1.70401 | 0 | 1.552609 | 0 |
| 1026.049 | 1.703426 | 0 | 1.551782 | 0 |
| 1029.451 | 1.702842 | 0 | 1.550954 | 0 |
| 1032.854 | 1.702259 | 0 | 1.550123 | 0 |
| 1036.257 | 1.701677 | 0 | 1.549291 | 0 |
| 1039.661 | 1.701095 | 0 | 1.548457 | 0 |
| 1043.065 | 1.700514 | 0 | 1.547621 | 0 |
| 1046.47 | 1.699933 | 0 | 1.546784 | 0 |
| 1049.874 | 1.699352 | 0 | 1.545944 | 0 |
| 1053.28 | 1.698772 | 0 | 1.545102 | 0 |
| 1056.685 | 1.698192 | 0 | 1.544259 | 0 |
| 1060.091 | 1.697613 | 0 | 1.543413 | 0 |
| 1063.497 | 1.697034 | 0 | 1.542566 | 0 |
| 1066.904 | 1.696455 | 0 | 1.541716 | 0 |
| 1070.311 | 1.695877 | 0 | 1.540864 | 0 |
| 1073.719 | 1.695298 | 0 | 1.540011 | 0 |
| 1077.126 | 1.69472 | 0 | 1.539155 | 0 |
| 1080.534 | 1.694142 | 0 | 1.538297 | 0 |
| 1083.943 | 1.693564 | 0 | 1.537437 | 0 |
| 1087.352 | 1.692986 | 0 | 1.536575 | 0 |
| 1090.761 | 1.692409 | 0 | 1.535711 | 0 |
| 1094.171 | 1.691831 | 0 | 1.534844 | 0 |
| 1097.581 | 1.691254 | 0 | 1.533976 | 0 |